\documentclass[10pt,a4paper]{article}
\usepackage{jheppub}
\usepackage{soul}
\usepackage{bm}
\usepackage{tikz}
\usepackage{bbm}
\usepackage{pifont}% http://ctan.org/pkg/pifont
\usepackage{braket}

\def\sec#1{{Sect.~\ref{#1}}}

\def\app#1{{Appendix~\ref{#1}}}

\allowdisplaybreaks

\newcommand{\be}{\begin{equation}}
\newcommand{\ee}{\end{equation}}
\newcommand{\bea}{\begin{eqnarray}}
\newcommand{\eea}{\end{eqnarray}}
\newcommand{\beq}{\begin{equation}}
\newcommand{\eeq}{\end{equation}}

\definecolor{mymagenta}{rgb}{1.0, 0.0, 0.56}

\definecolor{ForestGreen}{RGB}{34,139,34}

\makeatletter 
\gdef\@fpheader{}
\makeatother
\begin{document}

\title{Precision Tests in $\bm{b\to s\ell^+\ell^-}$ ($\bm{\ell=e,\mu}$) at FCC-ee}

\author[a,b]{Marzia Bordone,}
\emailAdd{marzia.bordone@cern.ch}

\author[b]{Claudia Cornella,}
\emailAdd{claudia.cornella@cern.ch}

\author[b]{Joe Davighi}
\emailAdd{joseph.davighi@cern.ch}

\affiliation[a]{Physik-Institut, Universit\"at Z\"urich, CH-8057 Z\"urich, Switzerland}

\affiliation[b]{Theoretical Physics Department, CERN, 1211 Geneva 23, Switzerland}

\date{\today}

\preprint{CERN-TH-2025-065, ZU-TH 22/25}

\abstract{The rare semi-leptonic decays $B\to K^\ast \ell^+\ell^-$, with $\ell=e, \mu$, are highly sensitive to new physics (NP) due to their suppression in the Standard Model (SM). Current LHCb measurements in the muon channel exhibit a significant tension with state-of-the-art SM theory predictions. The proposed tera-$Z$ run at FCC-ee provides a unique opportunity to untangle the origin of this tension by producing a very large sample of $B$-mesons in a clean $e^+e^-$  environment. We explore the expected precision of $B\to K^\ast \ell^+\ell^-$ ($\ell=e, \mu$) measurements at FCC-ee, complementing existing studies with $\tau^+\tau^-$ in the final state, and compare with HL-LHC projections.
For the case of muons in the final state, we show that HL-LHC and FCC-ee are expected to deliver a similar number of events,
while the latter performs much better in the case of final state electrons. Regardless of the lepton flavour, 
we expect the FCC-ee environment to be much cleaner than at HL-LHC, with subleading systematics. We also find that a significant reduction in theory uncertainties on the SM predictions is required to capitalize on the advantage going from HL-LHC to FCC-ee. 
We demonstrate the power of such measurements at FCC-ee to extract information on the long-distance contribution to these decays, and to reveal evidence for new physics even if no deviations are seen in electroweak precision tests.
}

\maketitle

\section{Introduction}

The leading role of flavour physics in exploring scenarios beyond the Standard Model (SM) makes it one of the most interesting sectors to probe at current and future colliders. 

Among the various processes, the rare flavour-changing neutral current $b\to s\ell^+\ell^-$ transitions are one of the most promising avenues to study the SM and its extensions. 
These transitions are forbidden at the tree-level in the SM, making it possible to probe a wide range of New Physics (NP) energy scales via precision measurements.
Of particular theoretical interest, broad classes of theories that solve the flavour puzzle close to the TeV scale, such as those based on flavour-deconstructed gauge interactions~\cite{Davighi:2023iks,Bordone:2017bld,Fuentes-Martin:2022xnb,Davighi:2022fer}, typically leave their leading phenomenological signatures in $b\to s\ell^+\ell^-$ transitions. But even more generally, these processes are excellent probes for {\em any} NP close to the TeV scale; for example, composite Higgs solutions to the hierarchy problem are, even when equipped with protective flavour symmetries, most strongly constrained by measurements of the helicity-suppressed decay $\bar{B}_s\to \mu^+\mu^-$~\cite{Glioti:2024hye,Stefanek:2024kds}.

One of the main channels where the $b\to s\ell^+\ell^-$ transition has been studied experimentally is the $B\to K^*\ell^+\ell^-$ decay. This mode offers a plethora of opportunities: from measurements of the branching fraction and its rich angular distribution \cite{Kruger:1999xa,Bobeth:2010wg,Bobeth:2012vn,Descotes-Genon:2013vna}, to tests of lepton flavour universality between different leptonic final states \cite{Hiller:2003js}. From the SM perspective, the predictions of lepton-flavour universality ratios are very precise, while the predictions for integrated branching fractions or angular observable are affected by non-factorisable, unknown long-distance effects due to charm re-scattering. This makes comparing SM predictions with experimental measurements hard.

Currently, LHCb is the experiment best suited to study $B\to K^*\ell^+\ell^-$ decays. Results for lepton flavour universality ratios between muon and electron final states are available with the full Run 2 datasets \cite{LHCb:2022qnv,LHCb:2022vje}, while angular analyses have so far used partial Run 2 data for muons \cite{LHCb:2020gog,LHCb:2020lmf} and the full Run 2 dataset for electrons~\cite{LHCb:2025pxz}.
Notably, the $B\to K^*\mu^+\mu^-$ branching fraction has only been measured with Run 1 data  \cite{LHCb:2016ykl}, and the di-electron mode has never been measured.
With the planned Upgrade II and a target dataset of $300\,\mathrm{fb}^{-1}$, LHCb aims to push the experimental precision of these observables toward the $\mathcal{O}(\%)$ level.

LHCb measurements currently exhibit significant tension with the SM predictions. This discrepancy becomes even more pronounced when combining various observables and decay modes, such as $B\to K\mu^+\mu^-$ \cite{LHCb:2016ykl}, $B_s\to\phi\mu^+\mu^-$ \cite{LHCb:2021zwz} or $\Lambda_b\to\Lambda\mu^+\mu^-$ \cite{LHCb:2018jna}. However, it remains unclear whether this tension originates from NP or from unaccounted-for non-factorizable effects. Attempts have been made in the literature to extract these effects directly from data \cite{LHCb:2023gel,LHCb:2023gpo,LHCb:2024onj,LHCb:2025pxz}, but a definitive answer is still lacking.

Another experimental possibility for studying flavour physics is a high-luminosity $e^+e^-$ machine running at the $Z$ pole, such as the proposed FCC-ee machine at CERN \cite{FCC:2018evy}. 
A tera-$Z$ run would produce many more heavy flavour mesons and tau leptons than is possible at a $B$ factory~\cite{Monteil:2021ith}, with significantly higher boost but still in a clean environment -- combining the strengths of $pp$ colliders and $B$ factories for performing precision flavour physics measurements.
Several studies have already highlighted the unprecedented improvements this would bring in several decay channels. Current $B$-physics studies focused on $b\to s\nu\bar\nu$, $b\to c\tau\bar\nu$ and $b\to s\tau\tau$ modes \cite{Kamenik:2017ghi,Li:2022tov,Ho:2022ipo,Amhis:2023mpj,Li:2020bvr,Amhis:2021cfy,Zuo:2023dzn,Miralles:2024iii,Altmannshofer:2025eor}, which are all marked out by having invisible particles in the final state, but many more have yet to be explored. As a result, NP studies of the FCC-ee flavour program have mostly focused on NP aligned with the third generation quarks and leptons~\cite{Allwicher:2023shc,Allwicher:2025bub}, where NP remains only weakly constrained and as is naturally predicted {\em e.g.} in the flavour deconstruction theories mentioned above. Complementary to this, several works have appreciated the significant power of electroweak precision measurements at tera-$Z$ to probe solutions to the flavour puzzle~\cite{Davighi:2023evx,Davighi:2023xqn,Capdevila:2024gki,Lizana:2024jby,Covone:2024elw}, revealing a powerful complementarity between electroweak and flavour measurements that can be achieved with tera-$Z$ level of precision. 

The scope of our work is to perform a pilot study of $b\to s\ell^+\ell^-$ transitions, focusing on $B\to K^*\ell^+\ell^-$ decays with $\ell =\mu\,, e$, for which we can infer the reconstruction efficiencies needed to obtain the FCC-ee projections from previous studies. We  explore the phenomenological consequences of their measurements at FCC-ee, also in combination with results from the High-Luminosity LHC (HL-LHC) and the FCC-ee electroweak precision program. Equipped with this study of the experimental prospects, we are therefore able to extend the picture for the Beyond the SM (BSM) potential of FCC-ee flavour measurements to the light leptons, complementing what is already known for the tauonic modes.

Independently of the origin of the currently observed discrepancies in $b\to s\mu^+\mu^-$ transitions, we demonstrate that FCC-ee has great potential to  provide information on the SM and to constrain NP in these transitions, especially for the di-electron final state. To illustrate this, together with the expected improvement from HL-LHC, 
we focus on the measurement of the binned $B\to K^*\ell^+\ell^-$ branching fractions, for both the $\ell = \mu$ and $\ell = e$ cases. 
We perform two analyses: in the first one, we compare the reach of HL-LHC and FCC-ee in extracting information on long-distance effects using experimental data, and in the second one, we study the potential to probe NP, both in simplified scenarios and in a specific set of UV models. In both cases, we provide estimates of the experimental sensitivity at HL-LHC and FCC-ee, and include projections for future theoretical uncertainties which may provide useful benchmarks also for further related studies.

This paper is organised as follows: in \sec{sec:2}, we review the current status of the theory predictions for $B\to K^*\ell^+\ell^-$ and $\bar B_s\to \mu^+\mu^+$, and define benchmarks for the theory predictions as well as for the expected sensitivity at HL-LHC and FCC-ee. In \sec{sec:3}, we present two studies. First, we study the current sensitivity of HL-LHC and FCC-ee to study long-distance contributions. We then explore the NP reach HL-LHC and FCC-ee, in combination with other datasets, like Electroweak Precision Observables (EWPOs) and Drell-Yan LHC data. We conclude in \sec{sec:4}.

\section{Current status and future projections}
\label{sec:2}

\subsection{Standard Model theory predictions for $b\to s\mu^+\mu^-$ mediated processes}
The $b\to s\ell^+\ell^-$ transitions are described in the SM by the following effective Lagrangian, defined at the $m_b$ scale:
 \begin{equation}
       \mathcal{L}_\mathrm{eff}(b\to s\ell^+\ell^-) = 
        \frac{4 G_F}{\sqrt{2}}  V_{tb}V^*_{ts}
  \sum_{i=1}^{10} C_i \mathcal{O}_i  ~ + ~\mathcal{L
}_{\rm QCD+QED}^{[N_f=5]} \,,
    \end{equation}
where $V_{ij}$ are CKM elements, and we assumed that $V_{cb}V^*_{cs} \approx - V_{tb}V^*_{ts}$. The relevant operators for our discussion are:
 \begin{align}
        \mathcal{O}_1 =& (\bar{s}^\alpha_{L}\gamma_\mu c^\beta_L)(\bar{c}^\beta_L\gamma^\mu b^\alpha_L)\,,   &  \mathcal{O}_2 =&(\bar{s}_{L}\gamma_\mu c_L)(\bar{c}_L\gamma^\mu b_L)\,,   \\
        \mathcal{O}_7=&\frac{e}{16\pi^2}m_b(\bar{s}_{L}\sigma^{\mu\nu} b_R)F_{\mu\nu}\,,    & \mathcal{O}_8=&\frac{g_s}{16\pi^2}m_b(\bar{s}_{L}\sigma^{\mu\nu}T^a b_R)G_{\mu\nu}^a\,,   \\
        \mathcal{O}_9=&\frac{e^2}{16\pi^2}(\bar{s}_{L}\gamma_\mu b_L)(\bar{\ell}\gamma^\mu \ell)\,,    &  \mathcal{O}_{10}=&\frac{e^2}{16\pi^2}(\bar{s}_{L}\gamma_\mu b_L)(\bar{\ell}\gamma^\mu\gamma_5 \ell)\,.
\end{align}
The complete basis, that also includes the subleading operators $\mathcal{O}_{3-6}$, can be found for example in
\cite{Altmannshofer:2008dz}.

We are interested in two $b\to s\ell^+\ell^-$-mediated processes: $B\to K^*\ell^+\ell^-$, with $\ell = \mu\,, e$, and $\bar B_s\to\mu^+\mu^-$. The description of $B\to K^*\ell^+\ell^-$ decays is rather involved. The hadronic $B(p)\to K^*(k)$ matrix elements are described by seven independent local form factors. For the vector current we use:
\begin{align}
    \bra{K^*(k, \eta)} \bar{s} \gamma^\mu b \ket{\bar{B}(p)}
        & = -\epsilon^{\mu\nu\rho\sigma} \eta^*_{\nu}(k)\, p_\rho\, k_\sigma \frac{2\, V(q^2)}{m_B + m_{K^*}}\,,
\label{eq:BVVFF}
\end{align}
while for the axial vector current:
\begin{align}
    &\bra{K^*(k, \eta)} \bar{s} \gamma^\mu \gamma_5 b \ket{\bar{B}(p)}= i \eta^*_{\nu} \times \nonumber\\
        &\times\bigg\lbrace
                2 m_{K^*} A_0(q^2) \frac{q^\mu q^\nu}{q^2}+ 16\frac{m_B  m_{K^*}^2}{\lambda} A_{12} \left[2 p^\mu q^\nu - \frac{M_B^2 - m_{K^*}^2 + q^2}{q^2} q^\mu q^\nu\right] \nonumber\\
        & \phantom{=} 
                + (M_B + M_{K^*})\, A_1(q^2) \bigg[g^{\mu\nu} + \frac{2(M_B^2 + M_{K^*}^2 - q^2)}{\lambda} q^\mu q^\nu- \frac{2(M_B^2 - M_{K^*}^2 - q^2)}{\lambda} p^\mu q^\nu\bigg]\bigg\rbrace\,,
                \label{eq:BVAFF}
\end{align}
and, finally, for the tensor current:
\begin{align}
    \bra{K^*(k, \eta)} &\bar{s} \sigma^{\mu\nu} b \ket{\bar{B}(p)}
         = i
        \eta^*_{\alpha}\epsilon^{\mu\nu}\!{}_{\rho\sigma}\times \phantom{\left\{-\times\left[\left((p+k)^\rho-\frac{M_B^2-M_{K^*}^2}{q^2}q^\rho\right)g^{\alpha\sigma}+
        \frac{2}{q^2} p^\alpha p^\rho k^\sigma\right]T_1(q^2)\right.}\nonumber\\
        &\left\{-\left[\left((p+k)^\rho-\frac{M_B^2-M_{K^*}^2}{q^2}q^\rho\right)g^{\alpha\sigma}+
        \frac{2}{q^2} p^\alpha p^\rho k^\sigma\right]T_1(q^2)\right. \nonumber\\
        & -\left.\left(\frac{2}{q^2}p^\alpha p^\rho k^\sigma-\frac{M_B^2-M_{K^*}^2}{q^2}q^\rho
        g^{\alpha\sigma}\right) T_2(q^2)+\frac{2}{M_B^2-M_{K^*}^2}p^\alpha p^\rho k^\sigma T_3(q^2)\right\}\,.
        \label{eq:BVTFF}
\end{align}
In these expressions, $q = p-k$ is the momentum transfer, $\eta$ is the polarisation vector of the $K^*$ meson, and we use the K\"all\'en function $\lambda = m_B^4+m_{K^*}^4+q^4-2 m_B^2 m_{K^*}^2-2 q^2 m_B^2-2 q^2 m_{K^*}^2$. Local form factors are currently available from Lattice QCD (LQCD) calculations only at very high $q^2$ and in the narrow width approximation for the $K^*$ \cite{Horgan:2013hoa,Horgan:2015vla}. Efforts to understand finite width effects in hadronic final states are ongoing, e.g. for the $\rho$ meson \cite{Leskovec:2025gsw}, but no results are yet available for the $K^*$. For our predictions, we use the interpolation between the existing LQCD results \cite{Horgan:2013hoa,Horgan:2015vla} and Light-Cone Sum Rules results in \cite{Bharucha:2015bzk} (see also \cite{Gubernari:2018wyi,Gubernari:2022hxn}).

Additionally, long-distance contributions from charm re-scattering, induced by the operators 
$\mathcal{O}_{1-2}$, must be included. These contributions effectively shift the Wilson coefficient 
$C_9$, potentially introducing a $q^2$-dependence due to charm re-scattering. 
Determining whether $C_9$ exhibits such a dependence—an unmistakable signature of long-distance 
effects—is a crucial test for any theoretical description of charm re-scattering. 
Various approaches have attempted to model these effects 
\cite{Khodjamirian:2010vf,Gubernari:2020eft,Gubernari:2022hxn,Ciuchini:2015qxb,Ciuchini:2021smi,
Bordone:2024hui,Isidori:2024lng,Capdevila:2017ert}, yet no consensus has been reached on their 
magnitude. 
For simplicity, we adopt the approach in \cite{Bordone:2024hui}, which is based on dispersion relations. 
We emphasize that our findings are largely insensitive to this choice, since we are merely aiming to estimate the theoretical and experimental reach for this decay mode. Note that we also promote the Wilson coefficient $C_7$ to an effective one in order to encode non-factorisable corrections to the $B\to K^*\gamma$ amplitude calculated in \cite{Beneke:2001at}. The value of $C_7^\mathrm{eff}(\mu_b)$ is kept fixed throughout our analysis.

With these choices, we can then start defining observables in $B\to K^*\ell^+\ell^-$ decays. For the purpose of this work, we will just focus on the normalised, binned branching fraction, defined as
\begin{equation}
    \frac{d\mathcal{B}(B\to K^*\ell^+\ell^-)}{dq^2}[q^2_\mathrm{min},q^2_\mathrm{max}] = \frac{1}{q^2_\mathrm{max}-q^2_\mathrm{min}}\int_{q^2_\mathrm{min}}^{q^2_\mathrm{max}}dq^2\,\frac{d\mathcal{B}(B\to K^*\ell^+\ell^-)}{dq^2}\,.
\end{equation}
Expressions for the differential branching fractions can be found in \cite{Altmannshofer:2008dz,Bobeth:2012vn}.

The theoretical description of the $\bar{B}_s\to\mu^+\mu^-$ mode is simpler. In the Standard Model, the branching fraction for the $\bar B_s\to\mu^+\mu^-$ reads:
\begin{equation}
    \mathcal{B}(\bar{B}_s\to\mu^+\mu^-)= \frac{\tau_{B_s}}{16\pi^3} {\alpha_\mathrm{EM}^2 G_F^2} |V_{tb}V^*_{ts}|^2f_{B_s}^2 m_{B_s} m_\mu^2 \left(1- \frac{4m_\mu^2}{m_{B_s}^2}\right)^{1/2} |C_{10}|^2\,,
    \label{eq:Bsmumu}
\end{equation}
where $f_{B_s}$ is the $B_s$ decay constant.  This expression must be corrected by power-enhanced QED effects, which have been computed in~\cite{Beneke:2017vpq,Beneke:2019slt}. We employ the numerical results from \cite{Beneke:2019slt}, choosing the $N_f=2+1+1$ result for the $B_s$ decay constant \cite{Bazavov:2017lyh,Hughes:2017spc,ETM:2016nbo,Dowdall:2013tga,FlavourLatticeAveragingGroupFLAG:2024oxs}. The input parameters used in our analysis are listed in Tables~\ref{tab:inputsFit}–\ref{tab:inputCi} in Appendix~\ref{app:inputs}.

\subsection{Future projections for the theory predictions}
Starting from the current theory predictions, we consider two benchmark scenarios in which we project  expected improvements in theory uncertainties. Concerning the hadronic $B\to K^*$ local form factors, we expect LQCD -- once  they overcome the difficulties in including finite width effects for the $K^*$ --  to reduce current uncertainties by a factor of 2 to 5 in the coming years \cite{Belle-II:2018jsg}. Given the currently unresolved challenges, we consider the improvement by a factor of 5 only as a long-term projection. We assume that uncertainties from non-factorisable effects entering $C_7^\mathrm{eff}$ scale similarly to those in the form factors.

For $\bar B_s\to\mu^+\mu^-$, improvements  in $f_{B_s}$ would require estimating isospin-breaking QED corrections, which are currently not understood for $B$ mesons. Therefore, we refrain from making any projections for $f_{B_s}$. However, we do explore  future improvements in the uncertainty on $|V_{cb}|$. We fix its central value to the inclusive one \cite{Finauri:2023kte}, which is preferred by CKM unitarity triangle fits \cite{UTfit:2022hsi}, and study two benchmarks: a reduction in relative error to $1\%$ and to $0.5\%$,  the latter motivated by the $W^+W^-$ threshold analysis \cite{Marzocca:2024mkc}. 
% \jd{For either scenario, the projected theory uncertainty on $\mathcal{B}(\bar{B}_s\to\mu^+\mu^-)$ can be computed as
% \begin{equation}
%     \frac{\sigma_{\mathcal{B}}}{\mathcal{B}} = \sqrt{4.38\times 10^{-4} + 4.14 \frac{\sigma_{V_{cb}}}{V_{cb}}}
% \end{equation}
% }

In summary, we define the following two scenarios for theoretical improvements:
\begin{align}
\mathrm{P}_1:\, &\sigma_{F_i} \to \sigma_{F_i}/2\,, \nonumber\\
&\sigma_{C_7^\mathrm{eff}} \to \sigma_{C_7^\mathrm{eff}}/2\,, \nonumber\\
& \frac{\sigma_{V_{cb}}}{V_{cb}} \to 1\%\,, \label{eq:P1} \\
\mathrm{P}_2:\, &\sigma_{F_i} \to \sigma_{F_i}/5\,, \nonumber\\
&\sigma_{C_7^\mathrm{eff}} \to \sigma_{C_7^\mathrm{eff}}/5\,, \nonumber\\
& \frac{\sigma_{V_{cb}}}{V_{cb}} \to 0.5\% \label{eq:P2} \,,
\end{align}
where we keep theory correlations fixed to the current values.
Our results for the expected precision of the SM predictions of $B\to K^*\ell^+\ell^-$ under these scenarios are shown in Table~\ref{tab:predictions_theory}.
For $\mathcal{B}(\bar{B}_s\to\mu^+\mu^-)$ this results in the following projected theory uncertainties 
\begin{equation}
    \sigma_{\mathcal{B}(\bar{B}_s\to \mu^+\mu^-)} = 
    \begin{cases}
        &1.07\times 10^{-10}\, \qquad (\mathrm{P}_1)\, \\
        &0.85\times 10^{-10}\, \qquad (\mathrm{P}_2)\, .
    \end{cases}
\end{equation}

\begin{table}[t]
    \centering
    \begin{tabular}{l|c|c|c||c|c|c}
        Bins & Current$|_{\ell= \mu}$ & $\mathrm{P_1}|_{\ell= \mu}$ & $\mathrm{P_2}|_{\ell= \mu}$& Current$|_{\ell= e}$ & $\mathrm{P_1}|_{\ell= e}$ & $\mathrm{P_2}|_{\ell= e}$ \\ \hline
        $q^2 \in [1.1,2.5]\,\mathrm{GeV}^2$ & $0.135$ & $0.068$ & $0.026$ &$0.135$ & $0.068$ & $0.026$\\
        $q^2 \in [2.5,4]\,\mathrm{GeV}^2$ & $ 0.133$ & $0.066$ & $0.026$ & $ 0.133$ & $0.065$ & $0.026$\\
        $q^2 \in [4,6]\,\mathrm{GeV}^2$ & $0.124$ & $0.060$ & $0.024$ & $0.124$ & $0.062$ & $0.024$\\
        $q^2 \in [6,8]\,\mathrm{GeV}^2$  & $0.112$ & $0.056$ & $0.023$ & $0.111$ & $0.056$ & $0.023$\\
    \end{tabular}
    \caption{Relative uncertainties on the binned branching fractions of $B\to K^*\ell^+\ell^-$, for the current situation and the projections $\mathrm{P_1}$ and $\mathrm{P_2}$ as defined in Eqs.~(\ref{eq:P1})-(\ref{eq:P2}).}
    \label{tab:predictions_theory}
\end{table}

\subsection{Experimental projections for HL-LHC and FCC-ee} \label{sec:exp_proj}

We begin by discussing the experimental projections for HL-LHC. For the $B\to K^*\ell^+\ell^-$ decays, we are interested in estimating the number of events per bin, $\frac{d\mathcal{N}^\mathrm{HL}_{\ell\ell}}{dq^2}[q^2_\mathrm{min},q^2_\mathrm{max}]$, which in turn allows us to infer the statistical  precision.

We use the signal yields from the latest $R_{K^*}$ analysis \cite{LHCb:2022qnv,LHCb:2022vje}, rescaled by the HL-LHC luminosity, finding that, in the central $q^2$ region ($q^2\in [1.1,6]\,\mathrm{GeV}^2$), this corresponds to $58000$ and $14776$ expected events for muon and electron final states, respectively. Since the branching fraction is rather flat across the $q^2$ window we consider, we assume that the number of events in each bin scales with the bin width. This gives:
\begin{equation}
  \begin{gathered}  \frac{d\mathcal{N}^\mathrm{HL}_{ee}}{dq^2}[1.1,2.5] =\, 4219   \,,  \qquad
    \frac{d\mathcal{N}^\mathrm{HL}_{ee}}{dq^2}[2.5,4] =\, 4520   \,, \\ 
    \frac{d\mathcal{N}^\mathrm{HL}_{ee}}{dq^2}[4,6] =\frac{d\mathcal{N}^\mathrm{HL}_{ee}}{dq^2}[6,8]\, =\,  6027 \,,
\end{gathered}
\end{equation}
for electrons, and 
\begin{equation}
\begin{gathered}
    \frac{d\mathcal{N}^\mathrm{HL}_{\mu\mu}}{dq^2}[1.1,2.5] =\, 16571   \,, \qquad 
    \frac{d\mathcal{N}^\mathrm{HL}_{\mu\mu}}{dq^2}[2.5,4] =\, 17755   \,, \\ 
    \frac{d\mathcal{N}^\mathrm{HL}_{\mu\mu}}{dq^2}[4,6] =\frac{d\mathcal{N}^\mathrm{HL}_{\mu\mu}}{dq^2}[6,8]\, =\,  23674 \,,
\end{gathered}
\end{equation}
for muons. From these estimates, we derive the expected relative statistical uncertainties: 
\begin{equation}
\begin{gathered}
    \sigma^\mathrm{HL}_{ee}[1.1,2.5] =\, 0.015   \,, \qquad
    \sigma^\mathrm{HL}_{ee}[2.5,4] =\, 0.015   \,, \\ 
    \sigma^\mathrm{HL}_{ee}[4,6] =\sigma^\mathrm{HL}_{ee}[6,8]\, =\,  0.013 \,.
\end{gathered}
\end{equation}
and
\begin{equation}
\begin{gathered}
    \sigma^\mathrm{HL}_{\mu\mu}[1.1,2.5] =\, 0.008   \,, \qquad
    \sigma^\mathrm{HL}_{\mu\mu}[2.5,4] =\, 0.008   \,, \\ 
    \sigma^\mathrm{HL}_{\mu\mu}[4,6] =\sigma^\mathrm{HL}_{\mu\mu}[6,8]\, =\,  0.006\,.
\end{gathered}
\end{equation}
In all these estimates we assumed that the $q^2\in [6,8]\,\mathrm{GeV}^2$ bin behaves as the $q^2\in [4,6]\,\mathrm{GeV}^2$ due to the smoothness of the branching fraction and equal bin widths.

While the statistical relative uncertainties obtained above are similarly sized in the electron and muon cases, current LHCb analyses show that analyses of $B\to K^*e^+e^-$ suffer from larger uncertainties than for $B\to K^*\mu^+\mu^-$. Moreover, electron statistics are currently insufficient to measure observables in more than one bin. This stems from the challenges in the electron reconstruction, that suffers from  substantial energy loss caused by photon emission. Therefore, our HL-LHC projections for the electron mode are likely optimistic.
Beyond statistical precision, systematic uncertainties must be considered. Estimating them properly would require a full detector-level simulation of selection and reconstruction efficiencies at LHCb. For our purposes, we assume that statistical and systematic uncertainty will be of the same size. This hypothesis is based on the expectation that many sources of systematic uncertainties will be better understood or modeled with higher luminosity.

For $\mathcal{B}(\bar{B}_s\to\mu^+\mu^-)$, we follow a similar procedure. The combination of the measurement by LHCb \cite{LHCb:2021awg,LHCb:2021vsc} and CMS \cite{CMS:2022mgd} was performed in \cite{Greljo:2022jac}. To project for the end of the HL-LHC phase, we rescale the current LHCb uncertainty as
\begin{equation}
\sigma_\mathrm{LHCb}^\mathrm{HL} = \sigma_\mathrm{LHCb}^\mathrm{current}\sqrt{\frac{9\,\mathrm{fb}^{-1}}{300\,\mathrm{fb}^{-1}}}\approx 0.08 \times 10^{-9}\,.
\end{equation}
This assumes that also the systematic uncertainty will scale with luminosity. This is justified at least for the  dominant contribution, namely the fragmentation fraction ratio $f_s/f_d$, which is expected to improve with more data. To include CMS prospects, we assume that CMS and LHCb will reach a similar precision. This hypothesis is supported by the current situation, that sees Run 1 + Run 2 results from  LHCb \cite{LHCb:2021vsc,LHCb:2021awg} and CMS \cite{CMS:2022mgd} yielding a similar precision.
With these assumptions, we obtain
\begin{equation}
\sigma_\mathrm{LHCb+CMS}^\mathrm{HL} = \frac{1}{\sqrt{2}} \sigma_\mathrm{LHCb}^\mathrm{HL}\approx 0.06 \times 10^{-9}\,.
\end{equation}

We now turn to the FCC-ee projections. We estimate the number of signal candidates for the differential branching fraction in a given bin $q_\mathrm{min}^2\leq q^2 \leq q_\mathrm{max}^2$ as
\begin{equation}
\frac{d\mathcal{N}}{dq^2}[q_\mathrm{min}^2,q_\mathrm{max}^2] =  N_Z \cdot \mathcal{B}(Z \to b \bar{b}) \cdot 2f_B \cdot  \frac{d\mathcal{B}(B \to K^* \ell^+\ell^-)}{dq^2}[{q_\mathrm{min}^2},{q_\mathrm{max}^2}] \cdot \epsilon_\mathrm{reco}\,,
\end{equation}
where $N_Z$ is the number of $Z$ produced at FCC-ee,  $\mathcal{B}(Z \to b \bar{b})$ is the branching fraction of $Z$ decaying to a $b\bar b$ pair, $f_B$ the fragmentation fraction into $B$ hadrons and $\epsilon_\mathrm{reco}$ is the overall reconstruction efficiency.
For our numerical analysis, we use $N_Z = 6 \times 10^{12}$, $\mathcal{B}(Z \to b \bar{b})=0.1512$ and $f_B = 0.407$. The reconstruction efficiency $\epsilon_\mathrm{reco}$ generally depends on $q^2$ and  determining it requires a dedicated detector-level simulation of the specific signal mode and its backgrounds. In this work, we conservatively set it to 80\% for both the electron and muon final states. This estimate is based on preliminary studies for $B\to K^*\tau^+\tau^-$
\cite{Miralles:2024iii}.

Our estimates for the expected number of events per bin and the corresponding relative statistical uncertainties are in Table~\ref{tab:FCC projections}. To obtain the total number of events per bin, we use the measured binned branching fraction from~\cite{LHCb:2016ykl} for the muon case.  For the electron case, which is not measured, we build a benchmark by rescaling the current branching fraction analysis for the muons with the measured value of $R_{K^*}$. Note that for both electrons and muons these benchmarks are 
 conservative since current experimental measurements see a downward shift relative to SM expectations. This results in a lower number of expected events and thus larger statistical uncertainties. Regarding systematics, we expect  these to be subleading at the FCC-ee thanks to the high boost of $b$ hadrons, and the fact that both electrons and muons can be reconstructed much better than at LHCb. We therefore neglect  systematic uncertainties in this projection.

\begin{table}[t]
    \centering
    \begin{tabular}{l|c|c}
        Bin & $\mathcal{N}$  &  $\sigma_\mathrm{stat}$   \\ \hline
        $q^2 \in [1.1,2.5]\,\mathrm{GeV}^2$ & $19289$ & $0.007$ \\
        $q^2 \in [2.5,4]\,\mathrm{GeV}^2$ &  $19672$ & $0.007$\\
        $q^2 \in [4,6]\,\mathrm{GeV}^2$ & $20943$ & $0.007$\\
        $q^2 \in [6,8]\,\mathrm{GeV}^2$ & $25374$ & $0.006$\\
    \end{tabular}
    \caption{Projected signal yields for $B\to K^*\ell^+\ell^-$ at FCC-ee, assuming the current measured central values. We report the expected number of events per bin and the corresponding relative statistical uncertainties. These estimates apply to both $\ell=\mu\,,e$.}
    \label{tab:FCC projections}
\end{table}

FCC-ee can also measure $\mathcal{B}(\bar{B}_s \to \mu^+\mu^-)$. This measurement is expected to be statistics-limited, but essentially background-free~\cite{Monteil:2021ith}. The study presented in~\cite{Monteil:2021ith} (based on a private study by Donal Hill) estimates that FCC-ee could reconstruct $\mathcal{N}_{\bar{B}_s\to\mu\mu}^{\mathrm{FCC}} \approx 540$ candidates. Assuming a more-or-less background-free measurement, this leads to a precision of roughly 5\% on the branching ratio: 
\begin{equation}
    \sigma_{\mathrm{FCC}} = \frac{1}{\sqrt{\mathcal{N}_{\bar{B}_s\to\mu\mu}}} \mathcal{B}(\bar{B}_s \to \mu^+\mu^-)  \approx 0.16 \times 10^{-9} \, .  
\end{equation}
Although this is not as precise as the HL-LHC projections, FCC-ee still provides a clean and complementary measurement. When combining the HL-LHC and the FCC results, the resulting uncertainty is dominated by the HL-LHC measurement:
\begin{equation}
\sigma_{\mathrm{LHCb+CMS+FCC}}\approx 0.056\times 10^{-9}\, .
\end{equation}

\section{Exploring the Physics Potential}
\label{sec:3}
In this section, we perform two analyses. The first concerns what can be learned with HL-LHC and FCC-ee about the short- or long-distance nature of contributions to $C_9^\ell$, where we introduce the superscript $\ell$ to distinguish our results for muons and electrons. We then move to the second  question, namely what is the NP reach at HL-LHC and FCC-ee using $b \to s \ell^+ \ell^-$ ($\ell = e,\mu$) measurements. For both analyses, we minimise the $\chi^2$ function
\begin{equation}
    \chi^2(\vec{C}) = \sum_{i} \left(\mathcal{O}_i^{\text{th}}(\vec{C}) -\mathcal{O}_i^{\text{exp}}\right)\Sigma_{ij}^{-1} \left(\mathcal{O}_j^{\text{th}}(\vec{C}) -\mathcal{O}_j^{\text{exp}}\right), 
\end{equation}
where ${\mathcal{O}i}$ denotes the set of observables considered, and the total covariance matrix is $\Sigma = \Sigma{\mathrm{th}} + \Sigma{\mathrm{exp}}$. In this expression, we use the results from \sec{sec:2}, and we keep the theory correlations fixed to the current ones for both P$_1$ and P$_2$.
For all the numerical analyses below, we restrict ourselves to the bins in the $q^2 \in [2.5,8]\, \mathrm{GeV}^2$ region. This is to overcome certain numerical instabilities that we observed for $q^2 <2.5\, \mathrm{GeV}^2$. We stress that, since our goal is to assess the potential of FCC-ee, this choice does not affect our conclusions.

\subsection{Extracting information on long-distance effects from $B\to K^*e^+ e^-$ and $B\to K^*\mu^+ \mu^-$} \label{sec:SMfit}

\begin{figure}[t]
    \centering
    \includegraphics[width=0.45\linewidth]{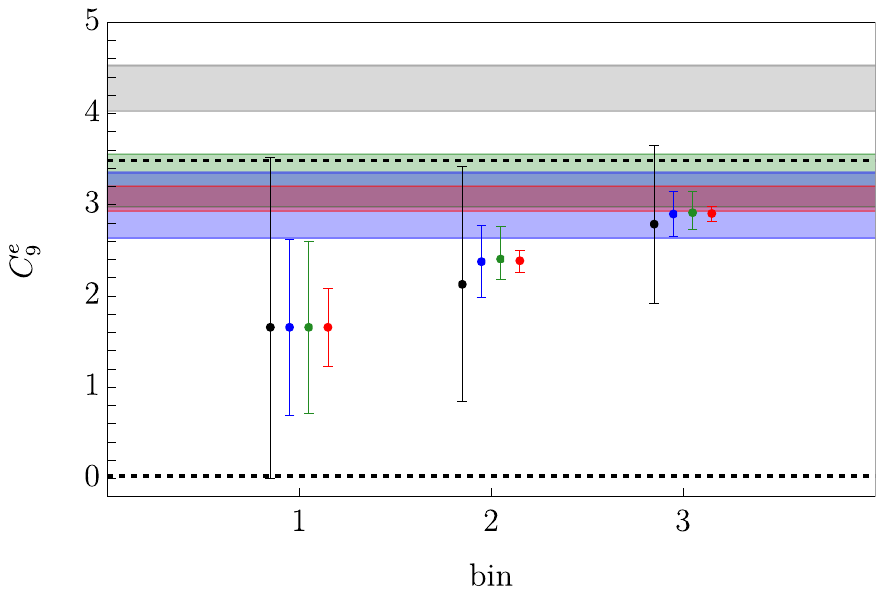}
    \includegraphics[width=0.45\linewidth]{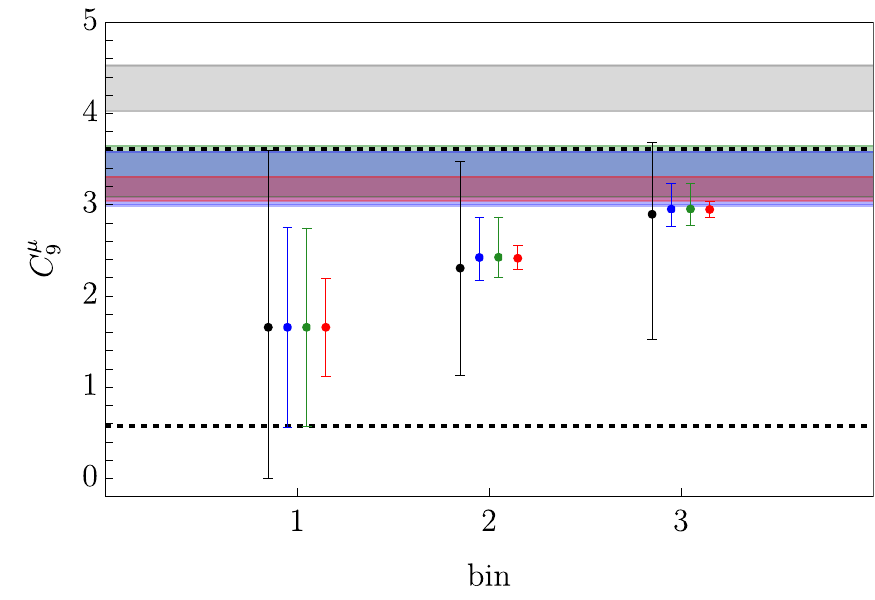}
    \caption{Extraction of $C_9^e$ and $C_9^\mu$ from the binned branching fraction. Black: current constraints; blue: HL-LHC projections; green: FCC-ee projections with $\mathrm{P}_1$ benchmark; red: FCC-ee projections with $\mathrm{P}_2$ benchmark.  The gray band denotes the SM prediction, and the dashed black lines show current LHCb constraints (where the constraint on 
$C_9^e$ is inferred using 
$R_{K^\ast}$ and muon data). Bins 1, 2, and 3 refer to the $q^2$ intervals $[2.5, 4]$, $[4,6]$, and $[6,8]~\mathrm{GeV}^2$ respectively. See text for details.
}
\label{fig:C9_extraction}
\end{figure}

In order to study possible long-distance effects in $B\to K^*\ell^+\ell^-$, 
we extract the value of $C_9^\ell$ from each bin in the 
region $q^2 \in [2.5,8] \, \mathrm{GeV}^2$  using the binned 
branching fraction measurement of $B\to K^*\mu^+\mu^-$ from Ref.~\cite{LHCb:2016ykl}. For the branching fraction of $B\to K^*e^+e^-$, we generate pseudo-experimental values by rescaling the muon-mode branching fractions with the measured value of $R_{K^{*}}$. We then compare these results against the hypothesis of $C_9^\ell$ being constant across the bins, testing which precision is needed to statistically assess if $C_9^\ell$ shows a $q^2$ dependence.

Our findings are summarized in Fig.~\ref{fig:C9_extraction}, 
where we contrast various scenarios against the gray band, 
which represents the SM prediction. 
The width of this band is dominated by scale uncertainties \cite{Isidori:2023unk}. 
The black binned points and the band enclosed by dashed black lines 
reflect the combination of current theoretical and experimental uncertainties. 
Notably, a constant $C_9^\ell$  provides an excellent fit to the current data for both $\ell = \mu, e$. 
The projections for the HL-LHC under the 
$\mathrm{P}_1$ benchmark for theoretical uncertainties are shown 
as blue points and bands in Fig.~\ref{fig:C9_extraction}. 
In this scenario, the fit quality deteriorates, but remains acceptable in the electron case, with the constant $C_9^e$ 
hypothesis yielding a p-value of 0.04. In contrast, the fit to constant $C_9^\mu$ has poor quality, with a p-value of $4\times 10^{-3}$.

For FCC-ee, we show results for two theory benchmarks: $\mathrm{P}_1$ (green) and $\mathrm{P}_2$ (red). In both cases, the assumption of a constant $C_9^\ell$ results in a poor fit, with p-values approaching zero for both lepton flavors.

Figure~\ref{fig:C9_extraction} highlights the crucial role of theoretical 
improvements, especially in exploiting the FCC-ee reach. Already at the HL-LHC, theoretical uncertainties are expected to dominate over experimental ones, both in the current scenario and in the $\mathrm{P}_1$ projection. A substantial gain in precision is achieved when combining FCC-ee data with the $\mathrm{P}_2$ benchmark. 
In this case, we repeat the fits under the assumption of SM-like experimental central values and find that the SM prediction for $C_9^\ell$ can be tested with 2\% precision. 
This level of precision is particularly striking given it is attained using a single observable. 
Combining analyses of angular observables in $B\to K^*\ell^+\ell^-$  and complementary channels like $B\to K\ell^+\ell^-$, as is currently done with LHCb measurements,
will further enhance sensitivity in $C_9^\ell$, potentially reaching the per-mille level. 
Across all projections, we observe that the third bin significantly drives the central value of the fit. Additionally, theory correlations play a crucial role in shaping the fit results. These correlations are expected to evolve with upcoming LQCD inputs. Consequently, the results shown in Figure~\ref{fig:C9_extraction} should be regarded as illustrative, reflecting the interplay between current theoretical and experimental expectations.

\subsection{New physics sensitivity}

We now study the potential of FCC-ee measurements of $B\to K^\ast \ell^+\ell^-$ to detect and/or constrain heavy new physics. To do this, we write the relevant low-energy Wilson coefficients as:
\begin{equation}
C_9^\ell = C_9^\mathrm{SM}+\Delta C_9^\ell\,, \quad \mathrm{and}\quad C_{10}^\ell = C_{10}^\mathrm{SM}+\Delta C_{10}^\ell\,,
\label{eq:DeltaC910}
\end{equation}
where $\Delta C_{9,10}^{\ell}$ denotes the NP contribution to these Weak Effective Theory (WET) operators introduced in \sec{sec:2}. With this parametrization, we extract the expected sensitivity of HL-LHC and FCC-ee to various new physics scenarios. We do so in three consecutive steps. First, we directly fit the shifts in the low-energy WET Wilson coefficients. Second, we translate these results into constraints on SM Effective Field Theory (SMEFT) coefficients \cite{Grzadkowski:2010es}, enabling a direct comparison with higher-energy probes such as EWPOs measured at LEP (and future prospects at FCC-ee), as well as Drell--Yan measurements at the LHC. At this stage we introduce a $U(2)^3$ flavour symmetry acting on the light quark families. With this choice, we study how this flavour symmetry connects low- and high-energy observables. Details of how we implement the $U(2)^3$ flavour symmetry are in \app{app:observables}. Finally, we consider a concrete UV model to illustrate yet further correlations with other observables that arise when going beyond EFT analyses. Details on the tree level matching between WET and SMEFT and on low-energy observables are in \app{app:observables}.

\subsubsection{WET} \label{sec:WET}

In this Section, we use our likelihood to estimate the sensitivity to shifts in the WET effective operators.
We do so under two different hypotheses for the future experimental central values: first, that they coincide with current LHCb measurements (which are in tension with SM predictions), and second, that they are SM-like. 

\begin{figure}
    \centering
        \includegraphics[width=0.45\linewidth]{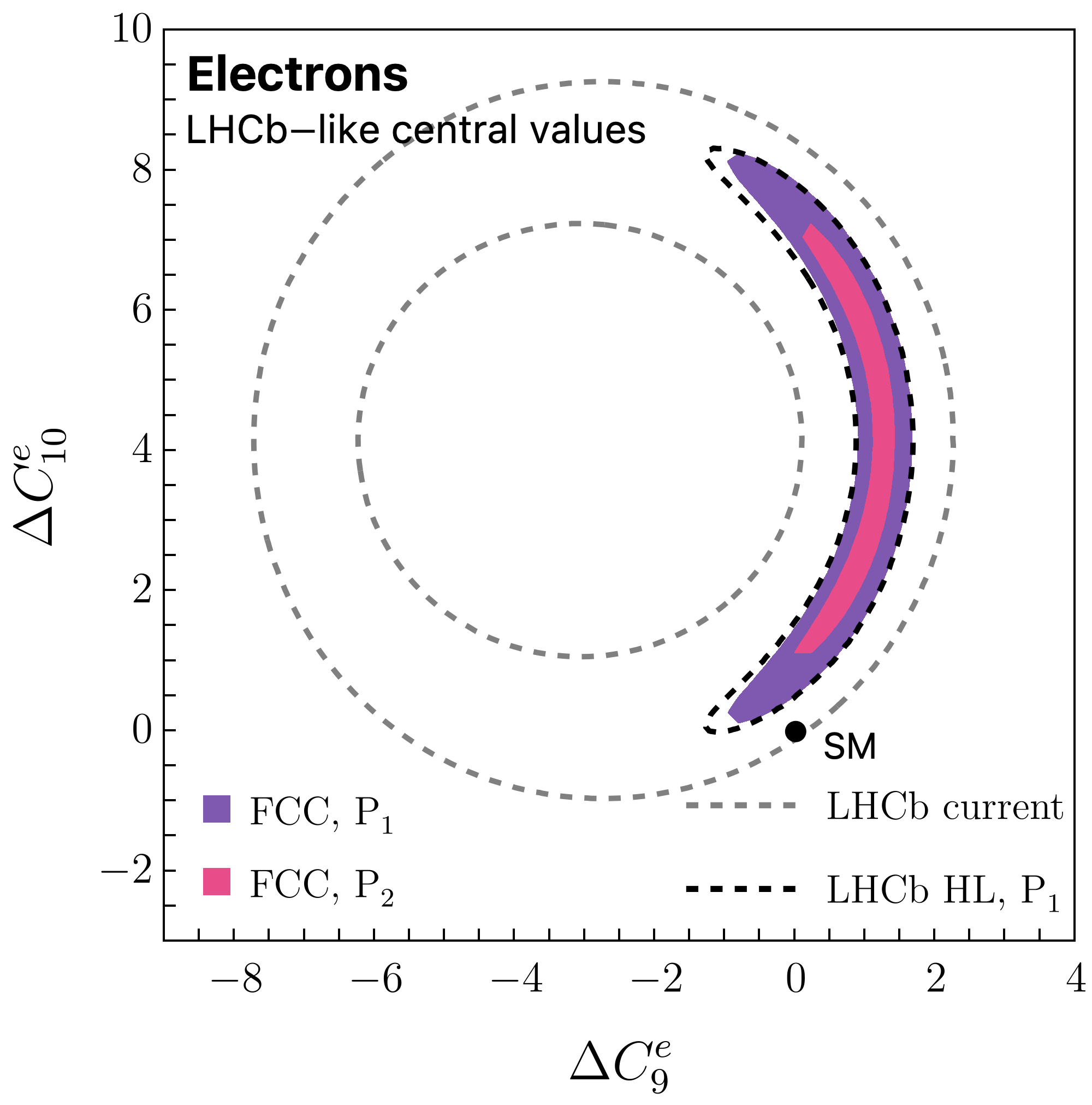}
        \,\,\,\includegraphics[width=0.45\linewidth]{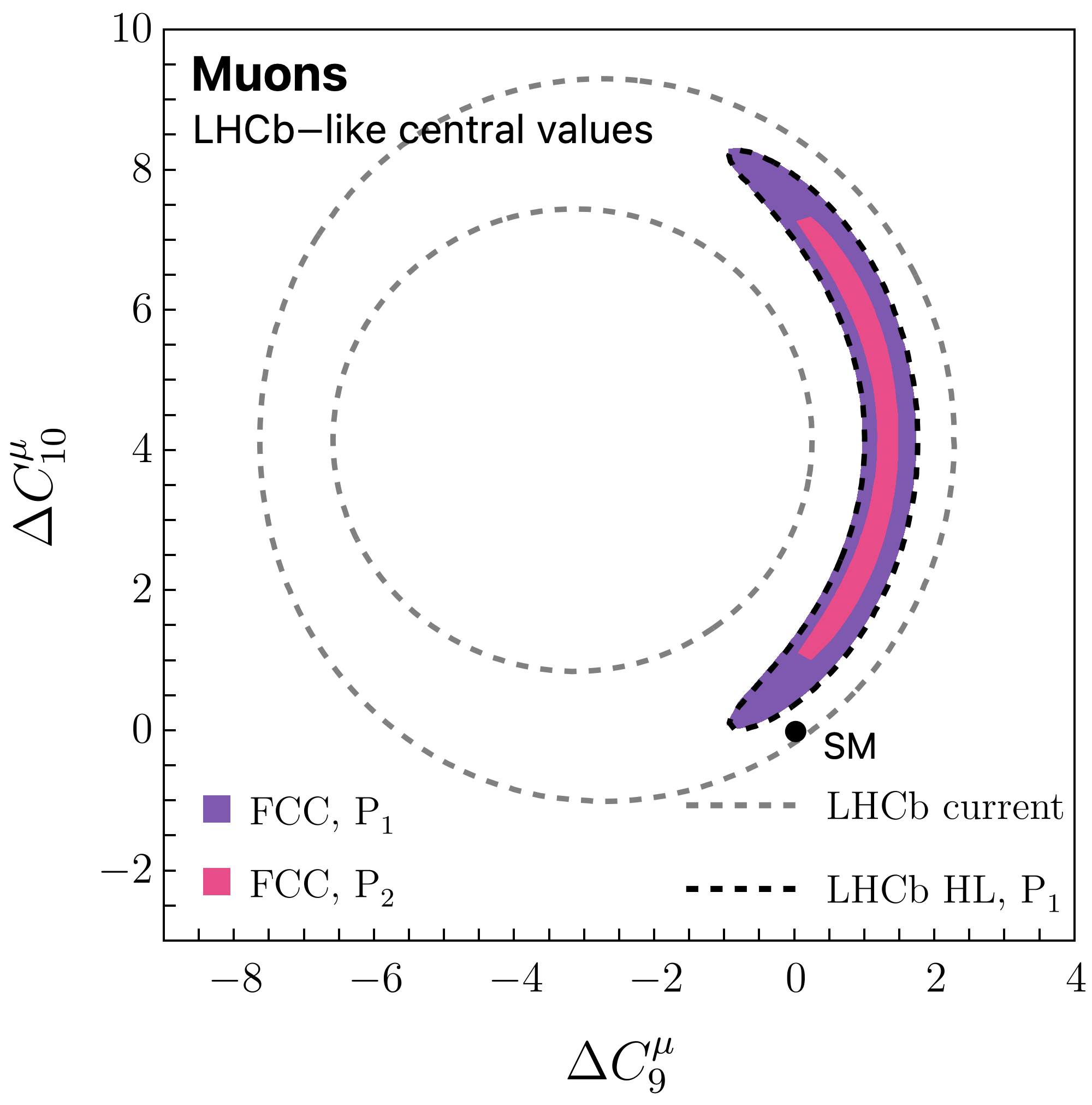}
    \caption{Projected fits to $\Delta C_9^{\mu,e}$ and $\Delta C_{10}^{\mu,e}$ from the binned branching ratio measurement of $B\to K^\ast\{ee,\mu\mu\}$, assuming future data follow current LHCb central values. All shaded regions denote $95\%$ CL.}
    \label{fig:c9c10}
\end{figure}

\begin{figure}
    \centering
        \includegraphics[width=.45\linewidth]{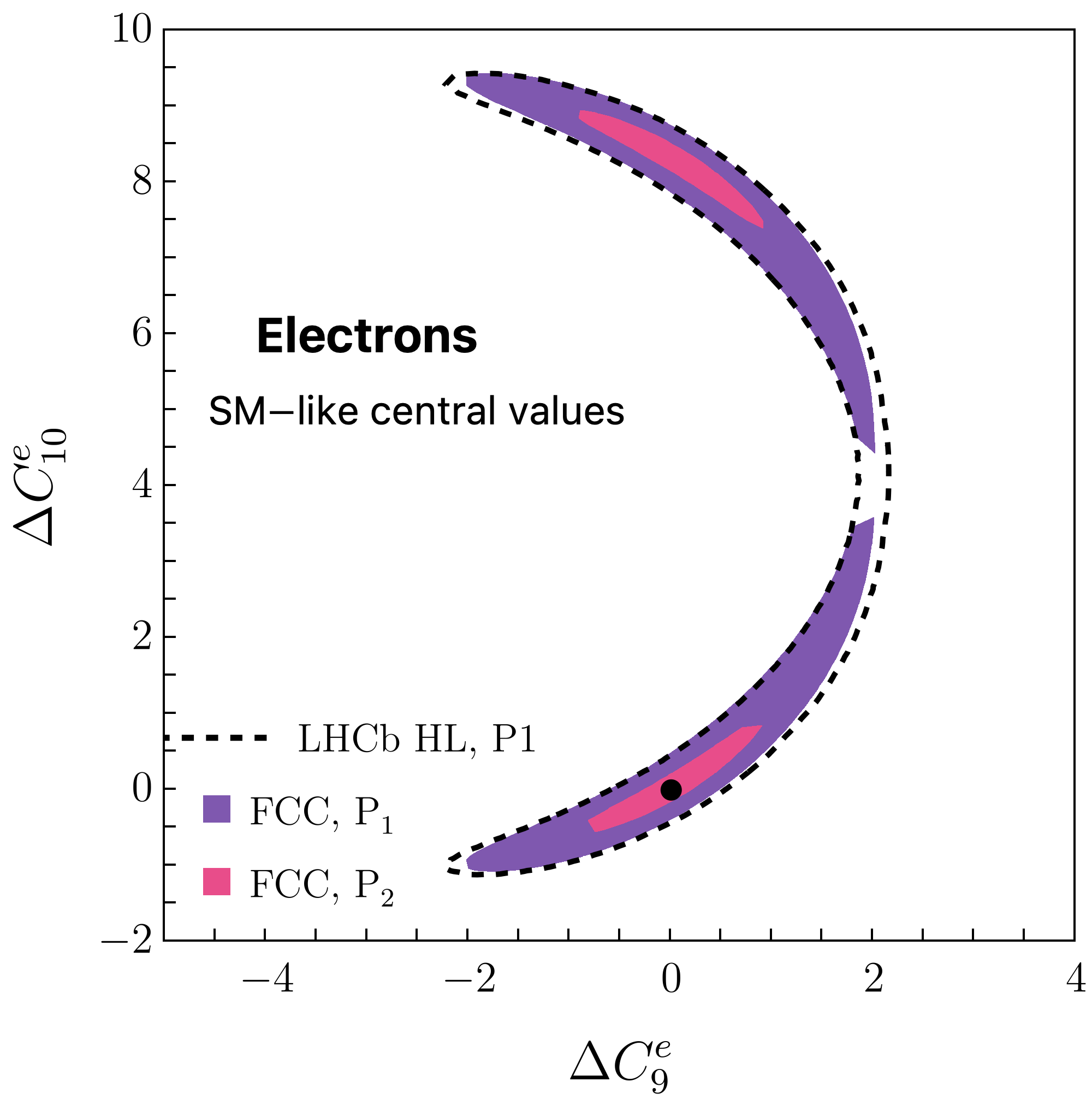}
        \,\,\,\includegraphics[width=.45\linewidth]{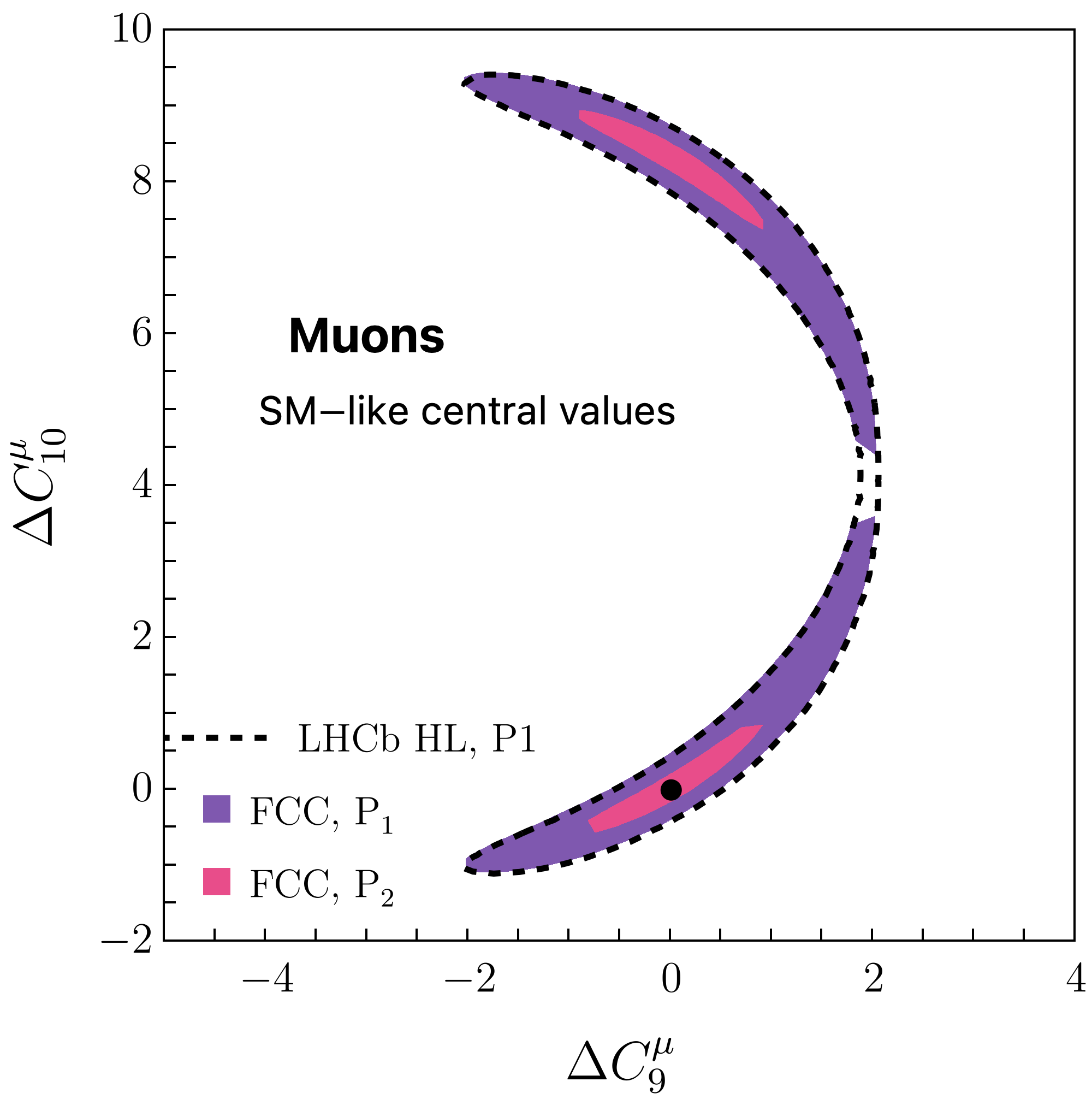}
    \caption{Projected fits to $\Delta C_9^{\mu,e}$ and $\Delta C_{10}^{\mu,e}$  from the binned branching ratio measurement of $B\to K^\ast\{ee,\mu\mu\}$, assuming future data have SM-like central values. All shaded regions denote $95\%$ CL.}
    \label{fig:c9c10_SMcentral}
\end{figure}

\begin{figure}
 \centering
    \includegraphics[width=.45\linewidth]{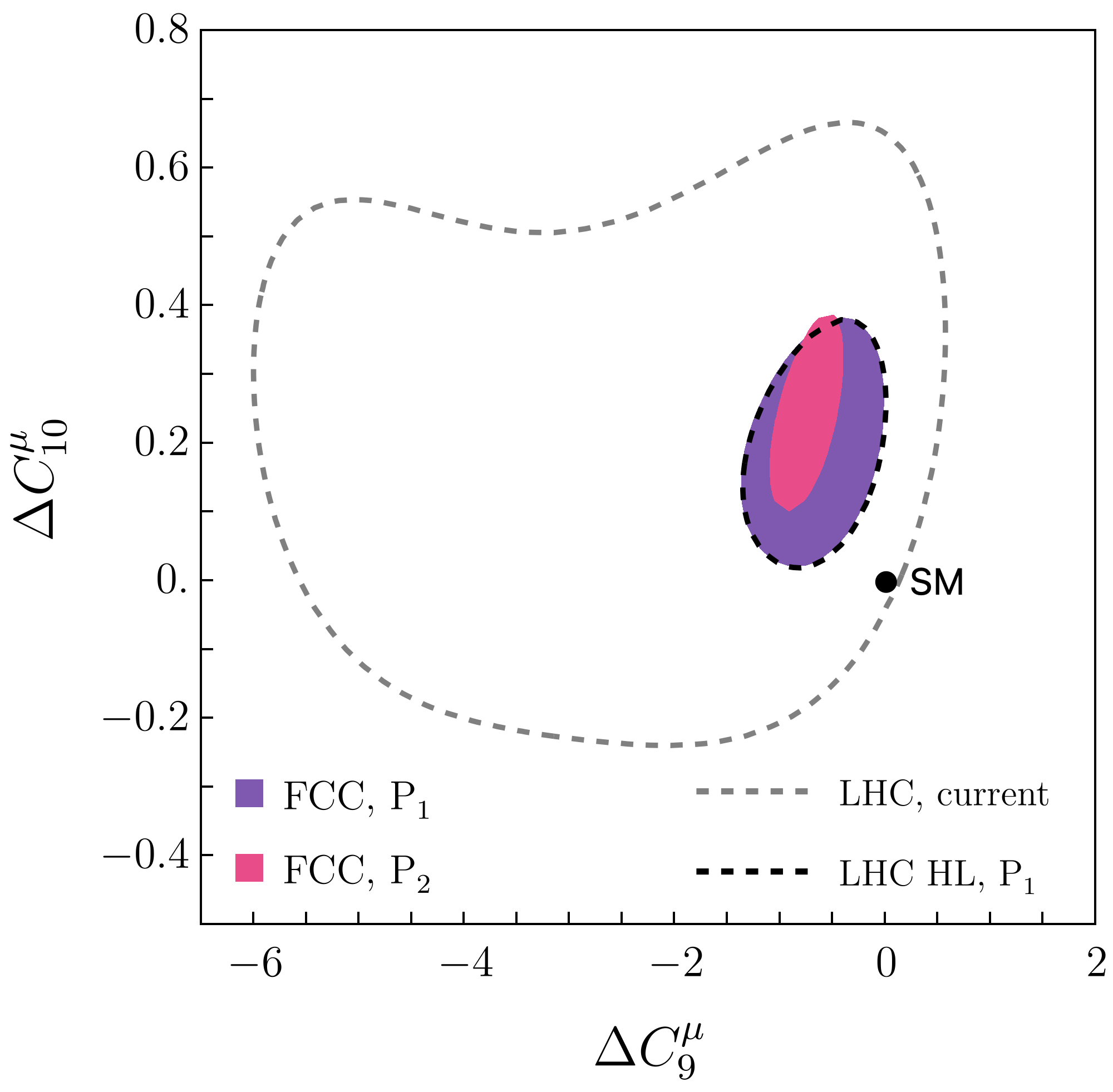}
    \includegraphics[width=.45\linewidth]{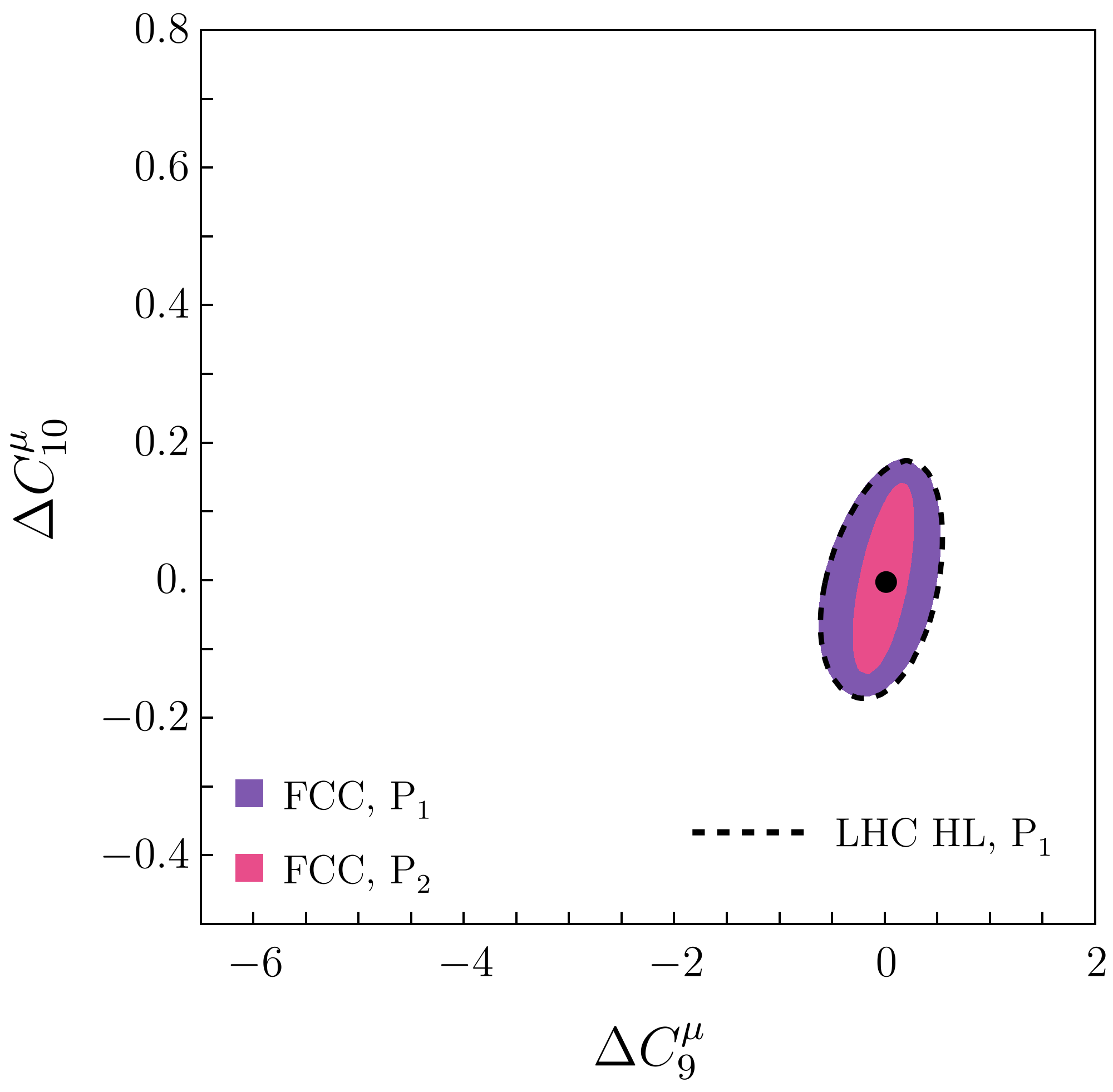}
    \caption{Projected improvements from current LHC results to HL-LHC and FCC-ee using the binned branching ratio of $B\to K^\ast\mu^+\mu^-$ and the branching ratio of $\bar{B}_s \to \mu^+\mu^-$, assuming future data have LHC-like (left) and SM-like (right) central values. 
    }
    \label{fig:c9c10_Bsmumu}
\end{figure}

The corresponding results are shown in  Figs.~\ref{fig:c9c10}--\ref{fig:c9c10_Bsmumu}, where we we plot the projected 95\% CL regions for several hypotheses and sets of observables. 
In Figs.~\ref{fig:c9c10} and~\ref{fig:c9c10_SMcentral} we focus on the binned branching fraction data for $B\to K^\ast \ell^+\ell^-$ alone, under two different hypotheses for the central values measured at future facilities: in Fig.~\ref{fig:c9c10} and Fig.~\ref{fig:c9c10_Bsmumu} (left) we assume future measurements follow the current LHCb central values (which, as mentioned, give slightly less statistical precision due to the observed under-fluctuation of the data),  while in Fig.~\ref{fig:c9c10_SMcentral} and Fig.~\ref{fig:c9c10_Bsmumu} (right)  we assume SM-like central values. 
In all cases we show the expected FCC-ee sensitivity under the benchmark theory uncertainty scenarios P$_1$ (\ref{eq:P1}) and P$_2$ (\ref{eq:P2}), alongside current LHC results and the projected HL-LHC reach under scenario P$_1$. The improvement from HL-LHC to FCC-ee is more pronounced in the electron channel, where one expects a substantial gain in statistics as well as the reduction in systematics. For the muon mode, on the other hand, the projected event yields at HL-LHC and FCC-ee are comparable, and the main advantage of FCC-ee lies in the expected superior control of systematics.
It is worth recalling that the current theoretical uncertainty on the SM predictions for the observables we study is already similar in size to the projected experimental uncertainty at HL-LHC. 
As a result, consistently with our findings from \sec{sec:SMfit}, the more conservative scenario P$_1$ leads to only modest improvements in new physics sensitivity when going from HL-LHC to FCC-ee, in both lepton channels.
In other words, the level of precision achievable at {\em both} HL-LHC and FCC-ee is so high that a significant reduction in theoretical uncertainties, as envisioned in scenario $\mathrm{P_2}$, will be necessary to fully exploit the superb FCC-ee statistics, which is particularly enhanced in the electron mode.

Finally, in Fig.~\ref{fig:c9c10_Bsmumu}, we focus on the muon channel and include the observable $\mathcal{B}(\bar B_s \to \mu^+\mu^-)$, which has been measured with impressive precision by LHCb \cite{LHCb:2021awg,LHCb:2021vsc}, CMS \cite{CMS:2022mgd}, and ATLAS \cite{ATLAS:2012qdc}. The theoretical prediction for this observable is lower than the experimental value and is dominated by the determination of CKM elements.
Since in absence of NP right-handed quark currents and of scalar currents $\mathcal{B}(\bar{B}_s\to\mu^+\mu^-)$ depends only on $C_{10}^\mu$, including this observable helps lifting the largely flat direction in $C_{10}$ evident in the $B\to K^\ast\mu\mu$-only plots.\footnote{Note that the corresponding exercise is not possible with the electron channel: due to the helicity suppression, the corresponding branching ratio with electrons is tiny, {\em viz.} $\mathcal{B}(\bar B_s \to e^+e^-)\sim 10^{-13}$ in the SM, and so even with future colliders it is likely that information on this process will be restricted to setting upper limits~\cite{LHCb:2020pcv}. } 
We use the projected sensitivity for $\mathcal{B}(\bar{B}_s\to \mu^+\mu^-)$ as described in \sec{sec:exp_proj}. These combined fits illustrate the significant potential of precision measurements in just this small selection of modes, {\em i.e.} $B\to K^\ast \mu^+\mu^-$ and $B_s \to \mu^+ \mu^-$, to exclude the SM point to high significance. However, they also underscore the pivotal role of  future improvements in theory predictions in assessing potential NP effects in these modes.

\subsubsection{SMEFT} \label{sec:SMEFT}

\begin{figure}
    \centering
    \includegraphics[width=.44\linewidth]{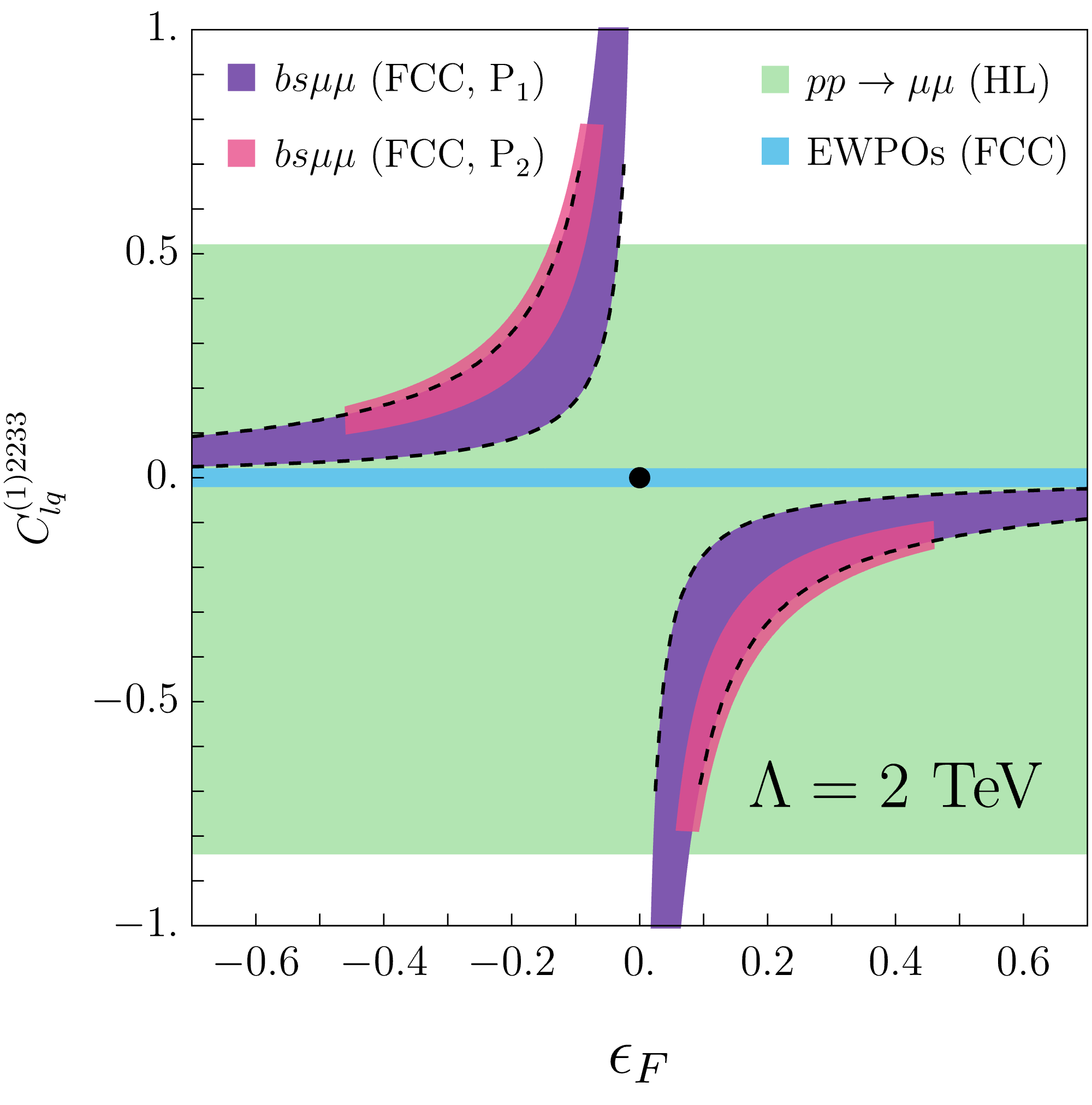}
    \,\,\,    \includegraphics[width=.44\linewidth]{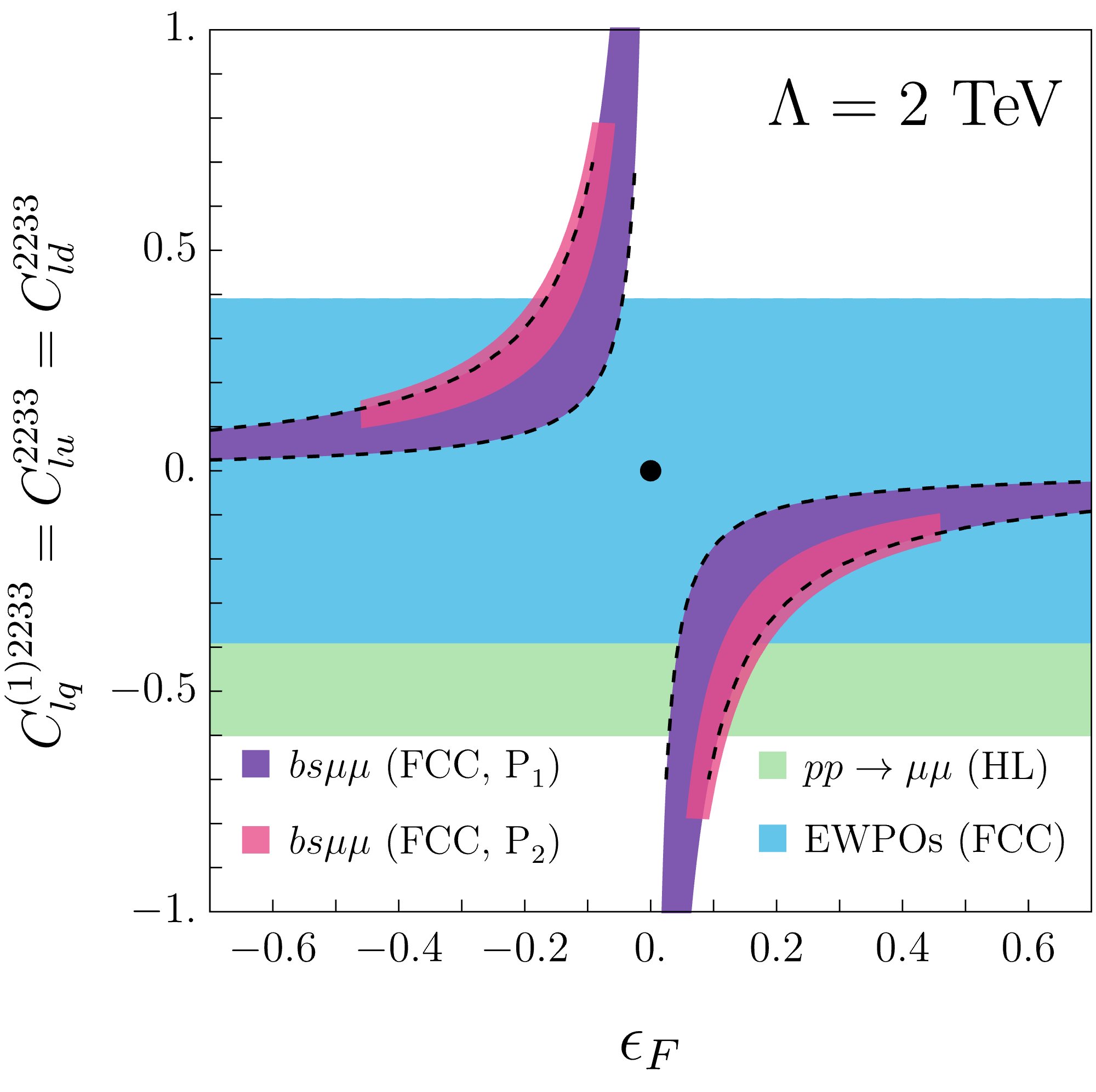}

    \includegraphics[width=.44\linewidth]{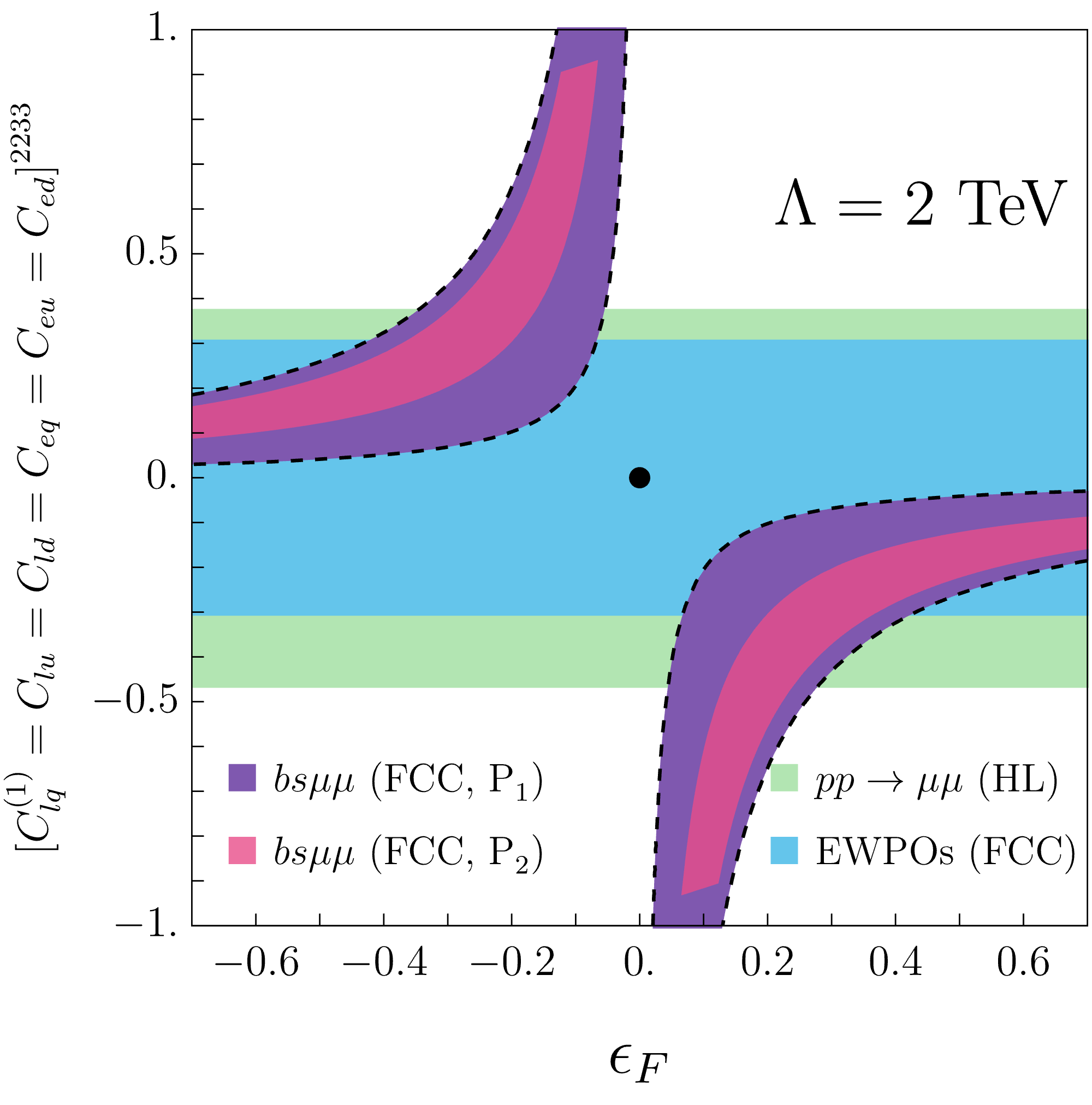}
    \caption{
    New physics sensitivity of measurements from HL-LHC and FCC-ee for the three SMEFT scenarios described in the main text. In pink/purple we plot the reach of our selection of flavour observables at FCC-ee (the dashed black lines indicate sensitivity from HL-LHC), alongside complementary constraints from EWPO measurements at tera-$Z$ (blue) and Drell--Yan data at the HL-LHC (green). Note that in the second plot the upper limit of the green bands nearly coincides with the upper limit of the blue bands and is therefore not visible.}
    \label{fig:SMEFT}
\end{figure}

We now move up in energy and interpret the shifts in the WET coefficients (\sec{sec:WET}) as arising from heavy new physics above the electroweak scale. 
We work within the SMEFT framework, assuming that new physics lies at a scale $\Lambda \sim $ few TeV and respects a $U(2)^3$ symmetry acting on the first two generations of quarks. 
The SMEFT description allows one to correlate the effects in the $b\to s\ell^+\ell^-$ ($\ell = \mu, e$) system with constraints coming from higher-energy datasets.  In the context of FCC-ee, it is particularly interesting to study the interplay with constraints from the electroweak precision measurements at the tera-$Z$ run.  Moreover, for scenarios where NP generates semi-leptonic operators at tree level but contributes to EWPOs only at one loop, one expects constraints from Drell--Yan data to provide complementary probes of the same parameter space. 
To study this interplay, we consider three benchmark scenarios, each defined by switching on a single SMEFT operator (or combination) at the matching scale $\Lambda = 2$ TeV:
\begin{itemize}
    \item  $C_{lq}^{(1)2233} \neq 0$: This scenario corresponds to new physics that couples left-handed muons to left-handed third-generation quarks. After rotation to the mass basis (see \app{app:observables}), this generates a coupling to the current $\overline{b}_L \gamma^\mu s_L$ + h.c., contributing to $\Delta C_9^\mu = - \Delta C_{10}^\mu$ at low energies. Such a  scenario could arise for instance by a heavy leptoquark coupled only to left-handed SM fermions.  
    \item $C_{lq}^{(1)2233}=C_{lu}^{2233}=C_{ld}^{2233}\neq 0$: Here the coupling to third-generation quarks is vector-like, while the coupling to muons remains purely left-handed. Such a vector-like coupling to quarks arises naturally in many models; for instance, if the new particle is a heavy gauge field, then a vector-like coupling is required to allow a renormalisable top Yukawa coupling (assuming that the Higgs is not charged under the new gauge group). At low energies, this setup again induces \( \Delta C_9^\mu = - \Delta C_{10}^\mu \). 
\item $C_{lq}^{(1)2233}=C_{lu}^{2233}=C_{ld}^{2233}=C_{eq}^{2233}=C_{eu}^{2233}=C_{ed}^{2233}\neq 0$: In this final case, new physics has vector-like couplings to both quarks and leptons. This scenario induces \( \Delta C_9^\mu \neq 0 \), with \( \Delta C_{10}^\mu = 0 \). Again, such a scenario would arise from a broad class of UV models featuring a spin-1 field coupled vectorially to the SM fermions (which can be consistently associated with breaking an extended gauge symmetry at a high scale). The UV model we study in \sec{sec:UV} is of this kind.  
\end{itemize}
In each case, we rotate to the mass basis via Eq.~\eqref{eq:Ld}. This introduces the parameter $\epsilon_F$, which quantifies the degree of down-alignment in units of $|V_{cb}|$.  Each scenario is then defined by two parameters: one Wilson coefficient and the alignment parameter $\epsilon_F$.

To assess the physics reach of FCC-ee and HL-LHC in these scenarios, we translate the results from \sec{sec:WET} into constraints on the SMEFT parameters. We then compare these bounds to two complementary sets of constraints: 
\begin{itemize}
    \item High $p_T$:
    The four-fermion SMEFT operators alter the tails of the $p p \to \mu^+ \mu^-$ distribution, enhancing the cross-section at high invariant mass. We use the \texttt{HighPT} package~\cite{Allwicher:2022mcg} to obtain a projection for the HL-LHC (assuming an integrated luminosity of 3 $\mathrm{ab}^{-1}$ and SM-like measurements) 
    \item Electroweak precision observables (EWPOs): even though our benchmark scenarios do not generate tree-level contributions to EWPOs, renormalization group effects lead to non-zero corrections to Z/W-pole observables at one loop.   We include these using the input scheme and observables of Ref.~\cite{Breso-Pla:2021qoe}, and assume SM-like future measurements with the expected FCC-ee precision as in~\cite{DeBlas:2019qco,DeBlas:2019qco,Blondel:2021ema,Bernardi:2022hny}. 
\end{itemize}
Our results are shown in Fig.~\ref{fig:SMEFT}. The flavour bounds (purple and pink) are overlaid with projections for HL-LHC (dashed black), FCC-ee EWPOs (blue), and HL-LHC high-$p_T$ (green). Several interesting features emerge.  In the left-handed scenario (upper-left panel in Fig.~\ref{fig:SMEFT}), EWPOs provide the most stringent constraints due to top-Yukawa enhanced RGE effects into Z-pole observables. In the vector-like scenarios instead, EWPO constraints are weaker, because the running into Z-pole observables is proportional to the $SU(2)_L$ gauge coupling.  Here,  constraints from EWPO at FCC-ee and Drell-Yan at the LHC are comparable in size. 
Notably, in these cases we see that the $b\to s\mu^+ \mu^+$ flavour observables provide a highly complementary (and indeed more precise)  constraint; notably, we see that these flavour measurements can reveal a significant tension with the SM even if EWPOs are measured to be consistent with the SM. 

These results demonstrate the power of precision flavour measurements to probe new physics that may otherwise evade detection in high-$p_T$ searches and EWPOs.  
They focus on the possibility of discovering new physics in the muon modes, since this is where a tension is observed. However, we emphasize that the bigger potential leap forward brought by FCC-ee in this domain is in the electron modes, where a big statistical improvement over HL-LHC is expected, in addition to the systematic improvement across both lepton channels. Comparable sensitivity to NP in $b\to s e^+e
^-$ is therefore expected at FCC-ee.

\subsubsection{Example UV model} \label{sec:UV}

\begin{figure}
    \centering
\includegraphics[width=.45\linewidth]{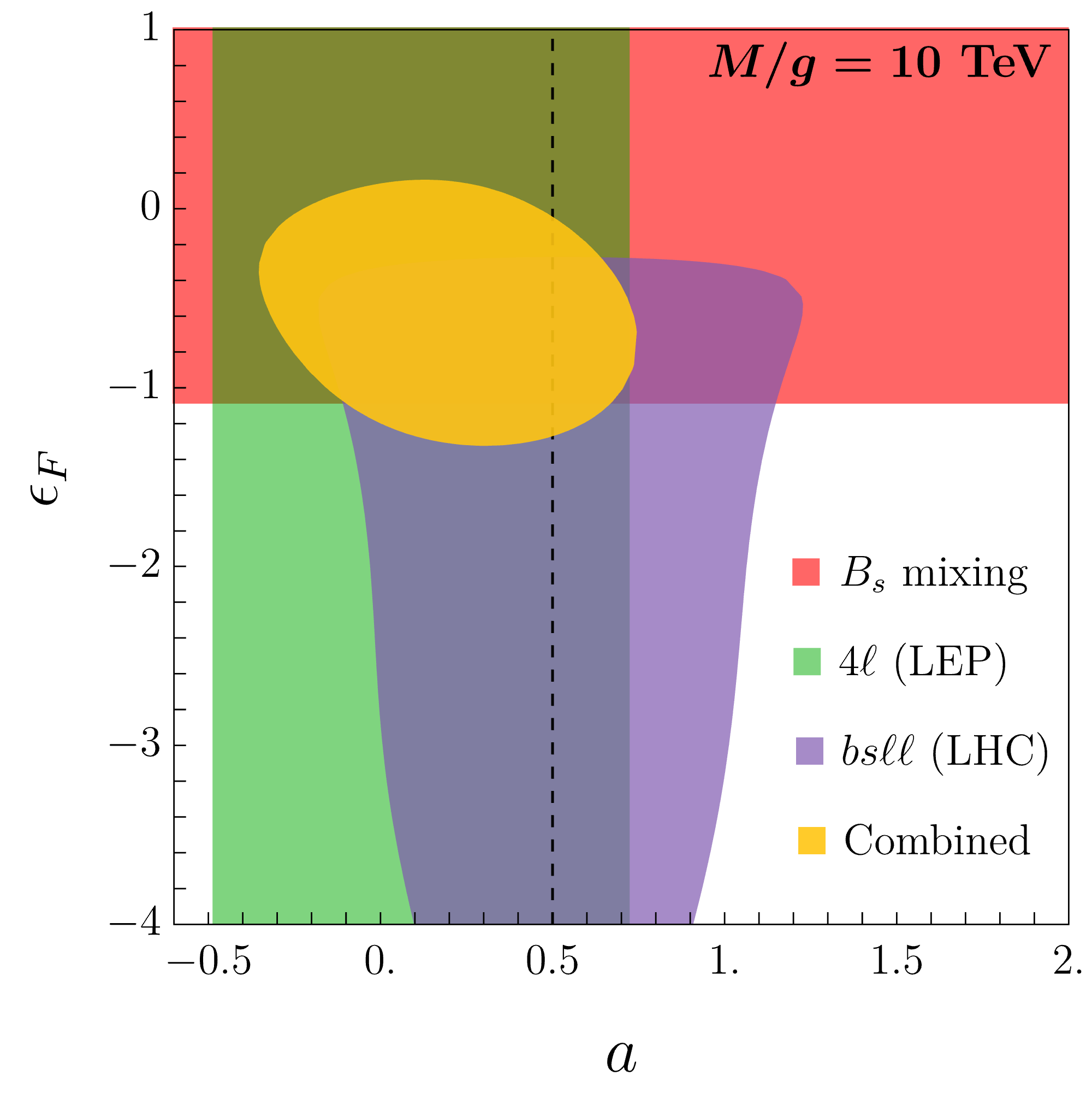}
\includegraphics[width=.45\linewidth]{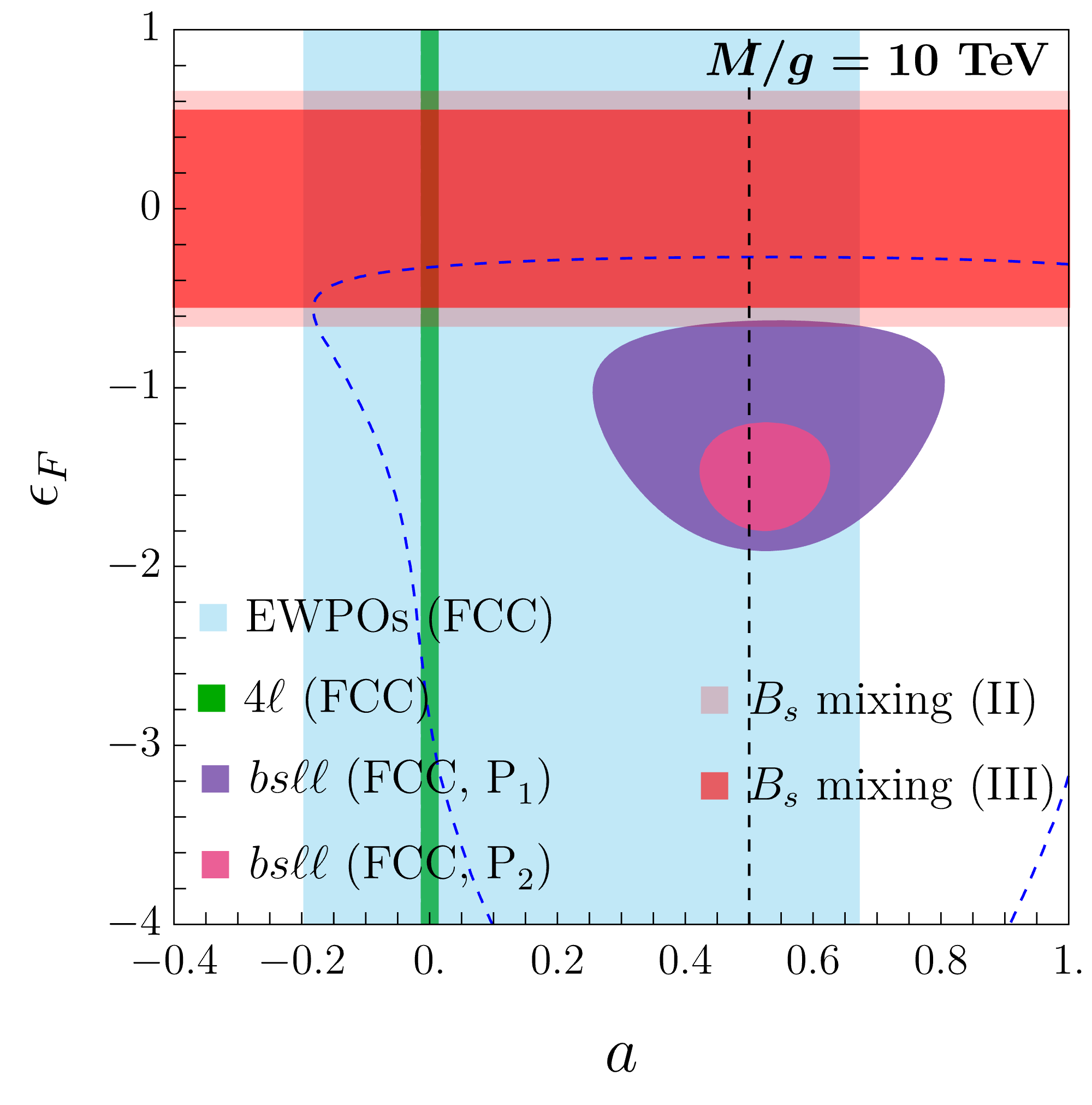}
    \caption{Constraints on the class of anomaly-free $Z^\prime$ models of Ref.~\cite{Allanach:2023uxz} (see text for details). Left: current constraints from $B_s$ mixing (red), LEP constraints on $e^+ e^-\to \ell^+\ell^-$ ($\ell = e, \mu$) above the $Z$ pole (green), and our selection of $b\to s\ell^+\ell^-$ observables (purple); we see reasonable compatibility of all these datasets, favouring some departure from LFU (indicated by the vertical black dashed line). 
    Right: FCC sensitivity to the same class of models, displaying orders of magnitude improvement coming from the $e^+e^-\to \ell^+\ell^-$ measurements (green) and flavour precision (purple and pink), together with an order-1 improvement in $B_s$ mixing~\cite{Charles:2020dfl} (red). The blue dashed line indicates the current $b \to s\ell^+\ell^-$ ($\ell = e,\mu$) constraints. }
    \label{fig:Zp}
\end{figure}

To illustrate how $b\to s\ell^+\ell^-$ measurements ($\ell = e,\mu$) at FCC-ee can probe explicit new physics models, we consider a class of anomaly-free $Z^\prime$ models introduced in~\cite{Allanach:2023uxz} as a possible explanation to the current tensions observed in $b\to s \mu\mu$ transitions. These models are based on gauging the family non-universal $U(1)_X$ symmetry
\begin{equation}
    X=B_3 - aL_1-(1-a)L_2 \,, 
\end{equation}
where $B_3$ denotes baryon number of the third generation and $L_{1,2}$ are the lepton numbers of the first and second generations. The gauge coupling $g_X$ is normalized such that the third-generation quark fields $q_3, u_3, d_3$ all carry unit charge, while $L_1, e_1$ have charge $-3a$ and $L_2, e_2$ carry charge $-3(1-a)$.
As for any $Z^\prime$ model, the couplings to third-generation up-type quarks are necessarily vector-like, to allow for a non-zero top Yukawa coupling at the renormalisable level\footnote{These models forbid the Yukawa couplings $y^u_{i3}$ and $y^d_{i3}$, for $i=1,2$, at the renormalisable level, and so it is a natural expectation in this framework that the CKM mixing angles $V_{ub}$ and $V_{cb}$ should be small, as accords with the observed values.}. As mentioned above, it is this property that leads to the cancellation of the $y_t$-running into EWPOs and thus to their weaker sensitivity to this family of models. 
By coupling only to $B_3$, the model enjoys an approximate $U(2)^3$ flavour symmetry in the quark sector. Upon rotating to the mass basis, the flavour non-universality in quark couplings is essential to generating the flavour violating $b\to s$ coupling at tree-level, which is a necessary ingredient for explaining the tensions in $b\to s\mu^+\mu^-$.
The coupling to leptons is also vector-like, hence this model most closely matches the third SMEFT scenario considered in \sec{sec:SMEFT}, except for the fact that it features non-zero couplings to electrons. The latter is controlled by the parameter $a$.
For $a=0$, the $Z'$ does not couple to electrons, as per the `$B_3-L_2$ model' studied in~\cite{Allanach:2020kss,Allanach:2022iod,Allanach:2022blr,Allanach:2024nsa}. For $a\approx 1/2$ instead the model is approximately lepton flavour universal in its coupling to electrons and muons, predicting $R_{K^\ast}\approx 1$ as favoured by current measurements.

The $Z'$ generates tree-level contributions to $b \to s \ell^+\ell^-$ transitions via flavour-changing couplings in the quark sector, and to other observables via four-fermion operators involving quarks and leptons. In addition to the constraints considered previously, the $Z^\prime$ model generates 4-quark and 4-lepton dimension-6 SMEFT operators at tree-level. This introduces correlated effects in other observables beyond those captured within a pure SMEFT analysis in \sec{sec:SMEFT}. The most relevant constraints come from:
\begin{itemize}
    \item {$B_s-\bar{B}_s$ mixing:}
    the $Z'$ contributes at tree level to the operator $(\bar s_L \gamma^\mu b_L)^2$. This leads to the bound~\cite{Charles:2020dfl}:
    \begin{equation}
        \frac{g_X^2}{2}\, \epsilon_F^2 \left(\frac{4.5\,\text{TeV}}{M}\right)^2 \leq 0.12 \qquad (95\%~\text{CL}).
    \end{equation}
    We also include future projections from~\cite{Charles:2020dfl}, referred to as {Phase II} (after HL-LHC and Belle II) and {Phase III} (including improved $|V_{cb}|$ precision from FCC-ee).
    \item {Cross-section and asymmetry measurements in $e^+e^-\to \ell^+ \ell^-$ collisions above the $Z$-peak}: at tree level, the $Z^\prime$ generates four-lepton operators
    \begin{equation}
    Q_{\ell\ell}^{iijj}\sim (\overline{\ell}_i \gamma^\mu \ell_i) (\overline{\ell}_j \gamma_\mu \ell_j), \qquad Q_{\ell e}^{iijj}\sim (\overline{\ell}_i \gamma^\mu \ell_i) (\overline{e}_j \gamma_\mu e_j),
        \qquad Q_{e e}^{iijj}\sim (\overline{e}_i \gamma^\mu e_i) (\overline{e}_j \gamma_\mu e_j),
    \end{equation}
for $i,j=1,2$. The combinations with $i(j)=1$ and $j(i)=1$ or $2$ are strongly constrained by LEP II measurements of $e^+e^- \to e^+ e^-$ and $e^+e^- \to \mu^+ \mu^-$.  At FCC-ee, these observables will be measured with far greater precision. To estimate it, we follow Ref.~\cite{Greljo:2024ytg} and construct a simplified likelihood using only the ratios $R_\ell=\sigma(e^+e^-\to\ell^+\ell^-)/\sigma(e^+e^-\to \text{hadrons})$\footnote{We are grateful to Alessandro Valenti for assistance here.}, which we expect to dominate the constraints.  
\end{itemize}
The constraints are summarised in Fig.~\ref{fig:Zp}, which shows the allowed region in the $(a, \epsilon_F)$ plane for $g_X = 1$ and $M = 10$ TeV. The left panel displays the current bounds, while the right panel illustrates the future FCC-ee sensitivity\footnote{A comprehensive global analysis of these models, incorporating a broader set of observables than the subset considered here, is presented in~\cite{Allanach:2023uxz}. Prospects at future colliders for a similar class of $Z^\prime$ models were studied very recently in~\cite{Allanach:2025qhd}. }.
We omit LEP constraints from $Z$ and $W$-pole observables, as these are only shifted at one-loop and are therefore significantly weaker than those shown.
Note that the region favoured by $b\to s\ell^+\ell^-$ data includes only our selected subset of observables: $\mathcal{B}(B \to K^\ast \mu^+\mu^-)$, $\mathcal{B}(B \to K^\ast e^+e^-)$ (which, as we mentioned above, is derived from the current measurement of $R_{K^\ast}$), and $\mathcal{B}(\bar{B}_s\to \mu^+\mu^-)$. Including additional correlated observables would significantly tighten this region. 
Acknowledging this caveat, we observe good compatibility among all these datasets, with LEP data favouring a small coupling to electrons, and thus some degree of LFUV that is nonetheless compatible with the measured $R_{K^\ast}$. Minimizing the total $\Delta\chi^2$ yields the orange 95\% CL region, which overlaps with all three regions going into the fit.

Turning to the future prospects (right plot), we see that at FCC this situation would dramatically shift.
 Projected sensitivities to $e^+e^-\to \ell^+\ell^-$ and flavour observables improve the constraints by several orders of magnitude, with an additional $\mathcal{O}(1)$ improvement in $B_s$ mixing~\cite{Charles:2020dfl}, as indicated by the phase II (III) bands.
The projected sensitivity for $e^+e^-\to \ell^+\ell^-$  imposes strong constraints on the electron coupling, as indicated by the dark green band. However, we caution that this projection assumes that $R_\ell$ off-peak measurements will be statistically dominated~\cite{Greljo:2024ytg}, and achieving such precision will require a substantial reduction of both systematic and theoretical uncertainties compared to the current state of the art.
Regarding flavour, if the current $R_{K^\ast}$ value persists, the improved precision on $B\to K^\ast \ell^+\ell^-$ ($\ell = \mu, e$) would strongly favour nearly equal NP effects in muons and electrons. Notably, the $a=0$ line would now be several sigma away from the region favoured by flavour measurements. 

Altogether, this combination of FCC-ee measurements has the potential to rule out the $Z^\prime$ explanation of the $b\to s\mu^+\mu^-$ anomaly, unless correlated deviations emerge elsewhere:  
the combined persistance of (i) an observed deviation in the muon channel, (ii) LFUV ratios measured to be approximately equal to 1, and (iii) SM-like measurements of $R_\ell$ at FCC-ee would become grossly incompatible with any $Z^\prime$ explanation.
Finally, we find that the sensitivity of Drell--Yan tails at HL-LHC to this model is much weaker than all the constraints shown, and is therefore not included in the figure.

\section{Conclusions}
\label{sec:4}
In this work, we have explored the potential of FCC-ee and HL-LHC for studying $b \to s \ell^+ \ell^-$ transitions, focusing on the $B \to K^* \ell^+ \ell^-$ decays with $\ell = \mu, e$. These processes provide a powerful avenue for probing both the Standard Model and possible extensions up to several TeV. By studying the binned branching fractions for these decays, we have shown that FCC-ee, in combination with HL-LHC data, can provide significant constraints on both long-distance effects and new physics contributions.
Particularly, the high reconstruction efficiency for the electron modes, and the very clean experimental environment, places the FCC-ee in a pivotal role to study the di-electron final states. As we see in all our analyses, the impressive expected experimental advancements that both HL-LHC and FCC-ee will deliver need to be supported by equally striking progress in the theory predictions. With our long-term projections for FCC-ee and theory advancements, we find that we will be able to test the presence of long-distance effects in $B\to K^*\ell^+\ell^-$ at the $\mathcal{O}(\%)$ level.

Furthermore, we study the new physics reach in various scenarios, with emphasis on the interplay with other low-energy observables such as the branching fraction of $\bar B_s\to\mu^+\mu^-$, as well as with high-energy data at the LHC and the FCC-ee projections for EWPOs. 
At the level of the weak effective theory, we find that combining the $B\to K^\ast \mu^+\mu^-$ projections with $\bar{B}_s \to \mu^+\mu^-$ is already sufficient to disfavour the SM with a high significance, if measurements were to to follow the central values observed by LHCb. At the level of the SM effective field theory, which encodes the effects of new physics above the EW scale, we demonstrate that, with approximate $U(2)^3$ flavour symmetries acting on light generations of quarks, there will be an excellent complementarity between these flavour probes and EWPOs. The latter are, of course, expected to be measured with extreme precision at FCC-ee. We highlight particular SMEFT scenarios, namely those in which the new physics couples ``vectorially'' to the top quark and not directly to the Higgs, for which the $b\to s \mu^+\mu^-$ measurements are more sensitive than the EWPOs. This shows that there are well-motivated BSM scenarios that could reveal themselves in precision flavour tests at FCC-ee, even if EWPOs are measured to be in agreement with the SM. Finally, this message is reinforced by considering a class of $Z^\prime$ models as an illustrative example, that currently offer a good description of the discrepancies in $b\to s \ell\ell$, but which can be covered completely by our precision flavour tests plus $e^+e^-\to \ell^+\ell^-$ cross-section measurements at FCC-ee.

Finally, we reiterate that this study lays the groundwork for future investigations of $b\to s\ell^+\ell^-$ transitions. A key next step is a full detector simulation and background reconstruction—similar to the approach taken in Ref.~\cite{Miralles:2024iii}—for $B\to K^*\ell^+\ell^-$, which will allow us to assess the FCC-ee sensitivity not only at low $q^2$ but also at high $q^2$. Even more intriguingly, these studies can be extended to include the $\ell = \tau$ case and its correlations with the light lepton modes. Notably, FCC-ee provides a unique environment for directly testing lepton flavour universality between the third and first/second generations in rare semi-leptonic $B$ decays, strengthening the leading role of FCC-ee for flavour physics.

\acknowledgments We thank St\'{e}phane Monteil and Mykhailo Yeresko for insightful discussion on the reconstruction efficiencies needed for this analysis. We thank Lukas Allwicher, Michele Selvaggi, and Alessandro Valenti for useful discussions. The work of MB was supported by the Swiss High Energy Physics initiative for the FCC (CHEF), with funding provided specifically by SERI and the University of Zurich.

\appendix
\section{Further inputs} \label{app:inputs}
The list of values for the input parameters of our analysis is in Tables~\ref{tab:inputsFit}-\ref{tab:inputCi}.

\begin{table}[t]
    \centering
    \begin{tabular}{c|c}
       Parameter & value \\ \hline 
        $\eta_{\mathrm{EW}}G_\mathrm{F}$ & $(1.1745\pm 0.0023) \times 10^{-5} \text{ GeV}^{-2}$ \\
         $ m_c $ & $ 1.68 \pm 0.20 ~ \text{GeV} $ \\
          $ m_b $ & $ 4.87 \pm 0.20 ~ \text{GeV} $ \\
        $ 1/\alpha_{\rm em}(m_b) $ & $ 133  $ \\
        $|V_{tb} V_{ts}^*| $ & $ 0.04185 \pm 0.00093 $ \\ 
            \end{tabular}
            \caption{Input parameter values used for our numerical analysis.}
            \label{tab:inputsFit}
        \end{table}

\begin{table}[t]
    \centering
    \begin{tabular}{l|c||l|c}
        Coefficient & value  &  Coefficient & value    \\ \hline
                $ C_1(\mu_b) $ & $ -0.291 \pm 0.009 $ &  $ C_6(\mu_b) $ &  $0.0012 \pm 0.0001$ \\
                $ C_2(\mu_b) $ & $ 1.010 \pm 0.001 $ &  $ C_7^{\mathrm{eff}}(\mu_b) $ & $ -0.450\pm 0.050 $ \\
                $ C_3(\mu_b) $ & $ -0.0062 \pm 0.0002 $ & $ C_8^{\mathrm{eff}}(\mu_b) $ & $ -0.1829 \pm 0.0006 $ \\
                $ C_4(\mu_b) $ & $ -0.0873 \pm 0.0010 $ &  $ C_9(\mu_b) $ & $ 4.273 \pm 0.251 $ \\
                $ C_5(\mu_b) $ & $ 0.0004 \pm 0.0010 $ &  $ C_{10}(\mu_b) $ & $ -4.166 \pm 0.033 $ \\
            \end{tabular}
            \caption{Input values used for the relevant Wilson coefficients; the operator basis we use is defined in~\cite{Altmannshofer:2008dz}.}
            \label{tab:inputCi}
        \end{table}

\section{Observables and EFT basis} \label{app:observables}
For the setup of the new physics analysis, we follow \cite{Allwicher:2023shc}. We start from the SMEFT \cite{Grzadkowski:2010es} above the electroweak scale, and we impose a $U(2)^3$ flavour symmetry acting on the light generations of quarks \cite{Barbieri:2011ci}. To describe  the processes we are interested in, we introduce the following effective Lagrangian at dimension 6
\begin{equation}
\mathcal{L} = \mathcal{L}_\mathrm{SM}^{(d=4)}+ \sum_i \mathcal{C}_iQ_i\,,
\end{equation}
where we consider the following effective operators:
\begin{align}
    Q^{(1) {prst}}_{\ell q} =&\, (\bar\ell_p\gamma_\mu\ell_r)(\bar q_s\gamma^\mu q_t)\,, & Q^{(3) {prst}}_{\ell q} =&\, (\bar\ell_p\gamma_\mu\tau^I\ell_r)(\bar q_s\gamma^\mu\tau^I q_t)\,, \\
    Q^{ {prst}}_{\ell u} =&\, (\bar\ell_p\gamma_\mu\ell_r)(\bar{u}_s\gamma^\mu u_t)\,, & Q^{(3) {prst}}_{\ell d} =&\, (\bar\ell_p\gamma_\mu\ell_r)(\bar {d}_s\gamma^\mu d_t)\,, \\
    Q^{(1) {pr}}_{H q} =&\, (H^\dagger i \overleftrightarrow{D}_\mu H)(\bar q_p\gamma^\mu q_r)\,, & Q^{(3) {pr}}_{H q} =&\, (H^\dagger i \overleftrightarrow{D}_\mu^I H)(\bar q_p\tau^I\gamma^\mu q_r)\,, \\
    Q^{(1) {prst}}_{q q} =&\, (\bar{q}_p\gamma_\mu q_r)(\bar q_s\gamma^\mu q_t)\,, & Q^{(3) {prst}}_{qq} =&\, (\bar{q}_p\gamma_\mu\tau^I q_r)(\bar q_s\gamma^\mu\tau^I q_t)\,,
\end{align}
where $\ell\,, q$ are the $SU(2)$ left handed lepton and quark doublets and $u$ and $d$ are right-handed singlets. In our notations, the Wilson coefficients $C_i$ have the dimension of the inverse of an energy squared.

Below the electroweak scale, we work with the Weak Effective Theory, the relevant part of which we introduced in the main text (\sec{sec:2}). To connect the SMEFT to the WET, we first introduce the rotation matrices in flavour space to account for the $U(2)^3$ symmetry (following~\cite{Fuentes-Martin:2019mun}):
\begin{equation}
 L_d \approx  \left(\begin{array}{c|c} 
 U_d &  {\begin{array}{c} 0  \\  \epsilon_F V_{cb} \end{array}}  \\ \hline
 \epsilon_F V_{td}  \quad   \epsilon_F V_{ts}  & 1
 \end{array} \right)\,, 
 \qquad U_d = 
 \left(\begin{array}{cc} 
 c_d & -s_d e^{i\alpha} \\
 s_d e^{-i\alpha}  & c_d \\
 \end{array}
 \right)\,,
 \label{eq:Ld}
 \end{equation}
with $s_d/c_d = |V_{td} /V_{ts}|$ and $\alpha_d ={\rm arg}
(V_{td}^*/V^*_{ts})$ and $\epsilon_F$ being a dimensionless parameter that measures the misalignment of the third generation with the up- or down-alignment. With this, we find the expressions for the various Wilson coefficients at low-energies.
For $b\to s\ell^+\ell^-$, the NP contributions $\Delta C_9^\ell$, as defined in Eq.~(\ref{eq:DeltaC910}), read
\begin{align}
    \Delta C_9^{\ell} &= \frac{\pi v^2}{\alpha_\mathrm{EM} V_{tb}V_{ts}^*} [L_d^\dagger]_{\alpha i} \Big[ C_{qe}^{[ij\ell\ell]} + C_{\ell q}^{(1)[\ell\ell ij]} + C_{\ell q}^{(3)[\ell\ell ij]} - \zeta \left(C_{Hq}^{(1)[ij]} + C_{Hq}^{(3)[ij]} \right) \Big] [L_d]_{jb} \,, \\
    \Delta C_{10}^{\ell} & = \frac{\pi v^2}{\alpha_\mathrm{EM} V_{tb}V_{ts}^*} [L_d^\dagger]_{\alpha i} \Big[ C_{qe}^{[ij\ell\ell]} - C_{\ell q}^{(1)[\ell\ell ij]} - C_{\ell q}^{(3)[\ell\ell ij]} + C_{Hq}^{(1)[ij]} + C_{Hq}^{(3)[ij]} \Big] [L_d]_{jb} \,,
\end{align}
where $\zeta = 1-4s_W^2$. We further investigate effects in $B_s$ mixing, that are described by
\begin{equation}
\mathcal{L}_{\Delta F = 2} = -C_{B_s}^1 (\bar s_L \gamma_\mu b_L)^2 \,, 
\end{equation}
where
\begin{align}
C_{B_s}^1 &=  - [L_d^\dagger]_{si} [L_d^\dagger]_{sk} \left( C_{qq}^{(1)[ijkl]} + C_{qq}^{(3)[ijkl]} \right) [L_d]_{jb} [L_d]_{lb} \,.
\end{align}

\bibliographystyle{JHEP}
\bibliography{refs}
%%%%%%%%%%%%%%%%%%%%%%%%%%%%%%%%%%%%%%%%%%%%%%%%%%%%%%%%%
\end{document}